\documentclass[prd,oneocolumn,notitlepage,aps,nofootinbib,nobibnotes,superscriptaddress,preprintnumbers]{revtex4-1}

\usepackage{color}%to be able to use colors
\usepackage{amsmath}%Recomended for mathematical documents (use of: eqref,...)
\usepackage{graphicx}
\usepackage{mathtools}
\usepackage{appendix}
\usepackage{amssymb}
\usepackage{verbatim}
\usepackage{ulem}
\usepackage{bm,bbm}
\usepackage[colorlinks=true,hyperfootnotes=true,citecolor=cyan]{hyperref}

\newcommand{\al}{\ensuremath\alpha}
\newcommand{\be}{\ensuremath\beta}
\newcommand{\ga}{\ensuremath\gamma}
\newcommand{\Ga}{\ensuremath\Gamma}
\newcommand{\del}{\ensuremath\delta}
\newcommand{\la}{\ensuremath\lambda}

\newcommand{\cR}{\ensuremath\mathcal{R}}
\newcommand{\cP}{\ensuremath\mathcal{P}}
\newcommand{\cQ}{\ensuremath\mathcal{Q}}
\newcommand{\cT}{\ensuremath\mathcal{T}}
\newcommand{\cS}{\ensuremath\mathcal{S}}
\newcommand{\cU}{\ensuremath\mathcal{U}}
\newcommand{\cB}{\ensuremath\mathcal{B}}
\newcommand{\cL}{\ensuremath\mathcal{L}}

\newcommand{\beq}{\begin{equation}}
\newcommand{\eeq}{\end{equation}}

\newcommand{\na}{\nabla}

\newcommand{\lr}[1]{\left(#1\right)}
\newcommand{\lrsq}[1]{\left[#1\right]}
\newcommand{\pa}[1]{\partial_{#1}}

\newcommand{\sq}{\bar{q}}
\newcommand{\tchi}{\tilde{\chi}}

\newcommand{\dd}{{\rm d}}

\newcommand{\bL}{\ensuremath\tilde{L}}
\newcommand{\bGa}{\ensuremath\bar{\Gamma}}

\newcommand{\cO}{\ensuremath\mathcal{O}}

\newcommand{\Upsilonh}{\hat{\Upsilon}}
\newcommand{\Deltah}{\hat{\Delta}}

\newcommand{\mpl}{M_{\rm Pl}}
\newcommand{\Ch}[3]{{\{^{\;{#1}}_{{#2}{#3}}\}}}

\begin{document}

\title{Instabilities in Metric-Affine Theories of Gravity}

\author{Jose Beltr\'an Jim\'enez}
\email[]{jose.beltran@usal.es}
\affiliation{Departamento de F\'isica Fundamental and IUFFyM, Universidad de Salamanca, E-37008 Salamanca, Spain.}

\author{Adri\`a Delhom}
\email[]{adria.delhom@uv.es}
\affiliation{Departament de F\'{i}sica Te\`{o}rica and IFIC, Centro Mixto Universitat de
Val\`{e}ncia - CSIC.\\
Universitat de Val\`{e}ncia, Burjassot-46100, Val\`{e}ncia, Spain}

\begin{abstract}
We discuss the presence of ghostly instabilities for metric-affine theories constructed with higher order curvature terms. We mainly focus on theories containing only the Ricci tensor and show the crucial role played by the projective symmetry. The pathological modes arise from the absence of a pure kinetic term for the projective mode and the non-minimal coupling of a 2-form field contained in the connection, and which can be related to the antisymmetric part of the metric in non-symmetric gravity theories. The couplings to matter are considered at length and cannot be used to render the theories stable. We discuss different procedures to avoid the ghosts by adding additional constraints. We finally argue how these pathologies are expected to be present in general metric-affine theories unless much care is taken in their construction.
\end{abstract}

\maketitle
\vspace{-0.5cm}

\small{\tableofcontents}

\section{Introduction}
Despite the impressive success of General Relativity (GR) in explaining gravitational phenomena in a wide range of scales \cite{Will:2014kxa} (including direct detection of gravitational waves \cite{Abbott:2016blz}), there is a logic in exploring gravitation beyond GR. On the observational side, the necessity of incorporating additional components to the universe inventory (e.g. dark matter and dark energy) as well as a satisfactory inflationary mechanism triggered a very active quest for explanations originating from a modification of gravity \cite{Clifton:2011jh,Joyce:2014kja}. On the theoretical side, the appearance of classical singularities such as the Big Bang or the black hole singularities, signal the breakdown of GR at high energies. This simply reflects the loss of perturbative unitarity of GR at around the Planck scale (in an optimistic scenario) that calls for a UV completion. Wishful thinking motivates exploring modified gravity scenarios where the classical evolution can be trusted before reaching the cut-off of the theory so that e.g. the Big Bang is replaced by a stable classical bounce or the central singularity of black holes is smoothed out by a wormhole or some other regular classical structures. This is the hope for instance in the Born-Infeld-like gravity theories (see e.g. \cite{BeltranJimenez:2017doy} and references therein.).

The intimate relation between GR and the geometry of the spacetime makes it natural to explore modifications to GR based on extending its geometrical framework, which in turn could be related to carefully accounting for the microscopic nature of spacetime itself. It is worth to emphasise that the very possibility of interpreting gravity on geometrical grounds roots in the most fundamental property of GR, namely: it describes an interacting massless spin-2 particle. Once the nature of the interaction has been established as such, one is naturally led to the conclusion that this particle must couple universally, thus giving rise to the equivalence principle and, consequently, to the very appealing relation of gravity to the geometry of spacetime. The precise geometrical framework where GR is to be interpreted is however conventional. We are most accustomed to regarding gravity as a manifestation of the spacetime curvature within a pseudo-Riemannian framework where the connection is devoid of any relevant role. However, this is just one possibility. In general, given a metric, the connection can be determined by its torsion (describing its antisymmetric part) and the non-metricity (measuring the failure of the connection in being metric-compatible). It is remarkable that the very same dynamics of GR can in fact be ascribed to these two objects in flat geometries \cite{Aldrovandi:2013wha,Maluf:2013gaa,Nester:1998mp,BeltranJimenez:2017tkd,BeltranJimenez:2019tjy,Jimenez:2019ghw}, i.e., geometries with a vanishing curvature. In this context, it seems a lawful act to grant the connection its rightful place in the geometrical scenario and construct theories where the metric and the connection are treated on equal footing. 

The generality of the metric-affine framework, where the connection introduces $D^3$ additional degrees of freedom (dof's) in $D$ dimensions, usually motivates the introduction of some restrictions either on the class of geometries or on the considered actions in order to work with manageable theories. Popular restrictions on the geometries are for instance the teleparallel framework, where the connection is imposed to be flat, or Weyl geometries and their generalisations, where the connection contains one additional vector field associated to the gauging of scale invariance. On the other hand, the connection can be left completely arbitrary and, then, make a judicious choice of the action so that the dynamics is substantially simplified. A paradigmatic example is the Einstein-Hilbert term that gives the same dynamics regardless the employed formalism because, in the metric-affine formulation, the connection is an auxiliary field whose equations of motion determine it to be the Levi-Civita connection up to an irrelevant projective mode. Another class of theories that has recently received attention is conformed by the so-called Ricci Based Gravity (RBG) where the action is constructed in terms of the Ricci tensor. Furthermore, most studies restrict it even further by imposing a projective symmetry so only the symmetric part of the Ricci tensor is allowed. This practical restriction turns out to be physically meaningful because it deprives the connection of any dynamics and, in fact, they are nothing but GR in disguise. 

A common argument in favour of general metric affine theories of gravity including higher order curvature terms is the manifest second order nature of the field equations. This property is sometimes claimed to avoid the presence of Ostrogradski instabilities that plague their corresponding metric formulation \cite{Woodard:2006nt,Woodard:2015zca}. However, this reasoning is flawed because what is sometimes not fully appreciated is the true origin of the pathologies. A very simple illustration of the incorrectness of this naive expectation is to notice that one can always reduce the order of some field equations by introducing suitable auxiliary fields. A perhaps better way of phrasing it would be to say that a criterion to avoid Ostrogradski instabilities is to ensure second order field equations on the constraint surface in phase space or, in other words, after having solved the constraints of the system and/or having integrated out all the auxiliary fields. In practice, this can lead to non-local actions that can obscure the number and nature of the physical dof's.

Thus, in order to properly diagnose the presence of Ostrgradski instabilities and/or other pathologies, a better strategy would be desired. The most direct way of tackling the issue would be to perform an exhaustive Hamiltonian analysis to identify the presence of constraints and work out the corresponding algebra. However, this can be technically difficult and, in many occasions, simpler approaches can be used to show the presence of pathologies. It is specially useful to check the breaking of gauge symmetries and the operators producing it\footnote{To give an explicit example, for a spin-1 field, breaking the $U(1)$ symmetry with non-derivative operators is safe, but derivative operators that break the gauge symmetry typically introduce ghostly modes.} as a manner to easily identify ghostly dof's. This approach was used in \cite{BeltranJimenez:2019acz} to show that generalized RBGs where the projective symmetry is explicitly broken generally contain additional ghostly dof's. These come in two forms: A spin-1 ghost arising from the absence of a pure kinetic term for the projective mode and Ostrogradski instabilities associated to non-minimal couplings of the additional 2-form field present in the theory. In \cite{Aoki:2019rvi}, it has been explored the important role played by the projective symmetry in the construction of healthy scalar-tensor theories. These studies put forward the relevance of the projective symmetry to avoid pathologies. In this work we will provide a more detailed analysis of RBGs and show the pathological character of non-projectively invariant RBG theories from different perspectives. Furthermore, we will argue how our findings naturally transcend to general metric-affine theories, thus making it clear that care must be taken when formulating theories with non-linear terms in the curvature even when formulated for a general affine connection. We will also illustrate how to evade these no-go results in different ways.

The paper is organised as follows. To make the paper as self-contained as possible, we start in Sec. \ref{sec:Generalities} by giving a complete and detailed introduction to the metric-affine formalism with all the necessary relations that we will use throughout the paper. In Sec. \ref{sec:RBG} we introduce the generalized RBG theories with the derivation of the field equations and the transformation to the Einstein frame. After introducing the general case, we briefly discuss on the projectively invariant theories in Sec. \ref{sec:PItheo} where we seize the opportunity to clarify some potential misconceptions. We then move to the main core of the paper in Sec. \ref{sec:GenRBG} where we study the non-projectively invariant theories. We discuss how the corresponding Einstein frame relates to the Non-symmetric Gravity Theory so that the results obtained for those theories are also applicable to generalized RBGs. We study a certain decoupling limit of the theory to show the appearance of ghosts that jeopardise the stability of the theories. We also show the problem by directly solving the connection equations. After diagnosing the sources of instabilities, in Sec. \ref{sec:Mattercoupling} we consider matter interactions that include direct couplings to the connection and show how they cannot, in general, stabilise the theories. In Sec. \ref{sec:GeoConst} we suggest several procedures that can lead to the avoidance of the ghosts, essentially by constraining the connection in different ways. We then consider a hybrid framework and show how they are even more prone to exhibit ghosts. We finally give our main conclusions in the discussions.

\section{Generalities of the metric-affine framework}\label{sec:Generalities}
Since we are going to deal with metric-affine gravity theories, it will be convenient to introduce the corresponding geometrical framework. A space-time is defined as a manifold endowed with a metric and an affine structures, determined by a metric $g_{\mu\nu}$ and a connection $\Gamma^\alpha{}_{\mu\nu}$ respectively. Unlike the metric formalism where the affine connection is assumed to be completely specified by the metric as the unique symmetric and metric-compatible connection, the metric-affine framework does not establish any relation {\it a priori} so the connection and the metric bear no relation between them, and both are accounted as fundamental fields. Despite the complete independence of the metric and the connection, it is often very useful to use the fact that given a metric there is a distinguished connection, namely the Levi-Civita connection, so the independent affine connection admits the following decomposition in three distinctive pieces:
\begin{align}\label{conndecomp}
\Ga^\al{}_{\mu\nu}=\bGa^\al{}_{\mu\nu}(g)+L^\al{}_{\mu\nu}(Q)+K^\al{}_{\mu\nu}(\cT).
\end{align}
These pieces are respectively called Christoffel symbols of $g_{\mu\nu}$, distortion tensor and contortion tensor and are given by:
\begin{align}
&\bGa^\al{}_{\mu\nu}(g)=\frac{1}{2}g^{\al\la}\lr{\pa{\mu}g_{\la\nu}+\pa{\nu}g_{\mu\la}-\pa{\la}g_{\mu\nu}}\;,\label{Crissymb}\\
&L^\al{}_{\mu\nu}(Q)=\frac{1}{2}\lr{Q^\al{}_{\mu\nu}-Q_{\mu\;\;\nu}^{\;\;\al}-Q_{\nu\;\;\mu}^{\;\;\al}}\;,\label{disformation}\\
&K^\al{}_{\mu\nu}(\cT)=\frac{1}{2}\lr{\cT^\al{}_{\mu\nu}+\cT_{\mu}{}^\al{}_\nu+\cT_\nu{}^\al{}_\mu},\label{contortion}
\end{align}
where $Q_{\al\mu\nu}\equiv\na_\al g_{\mu\nu}$ and $\cT^\al{}_{\mu\nu}\equiv2\Ga^\al{}_{[\mu\nu]}$ are the non-metricity and torsion tensors respectively. Notice that whereas the torsion is a property only of the affine connection, the non-metricity defines the relation between the metric and the symmetric part of the connection, and thus it is not a property of the affine connection alone. Notice also that all the non-tensorial behaviour of the connection is encoded in the Christoffel symbols, since the distortion and contortion pieces are indeed rank 3 tensors, since they can be regarded as the different of two connections. As a consequence, whereas one could choose given coordinates to make the connection, or the Christoffel symbols of $g_{\mu\nu}$ vanish at a point, this cannot be done for the contortion and distortion pieces. The curvature associate to the connection is given by the Riemann tensor, which reads
\beq
\cR^\al{}_{\be\mu\nu}(\Ga)\equiv\pa{\mu}\Ga^\al{}_{\nu\be}-\pa{\nu}\Ga^\al{}_{\mu\be}+\Ga^\al{}_{\mu\la}\Ga^{\la}{}_{\nu\be}-\Ga^\al{}_{\nu\la}\Ga^{\la}{}_{\mu\be}.
\eeq
For an arbitrary affine connection and in the presence of a metric, we can define the following three independent traces of the Riemann tensor:
\begin{align}
&\cR_{\mu\nu}(\Ga)=\cR^\al{}_{\mu\al\nu}(\Ga)\;,\label{Ricci}\\
&\cP^\mu{}_{\nu}(g,\Ga)=g^{\al\be}\cR^\mu{}_{\al\nu\be}(\Ga)\;,\label{coRicci}\\
&\cQ_{\mu\nu}(\Ga)=\cR^\al{}_{\al\mu\nu}(\Ga)\;,\label{Homothetic}
\end{align}
which are called the Ricci, co-Ricci and homothetic tensors respectively. The trace of the Ricci tensor defines the Ricci scalar $\cR(\Ga)=g^{\mu\nu}\cR_{\mu\nu}$ which also coincides with the trace of the co-Ricci tensor. Since the homothetic tensor is antisymmetric by construction, its trace vanishes. Notice that while the Ricci and the homothetic tensor exist in the absence of a metric, the co-Ricci tensor and the Ricci scalar do not. Another relevant property worth mentioning is the symmetries of a general Riemann. While the (covariant version of the) Riemann tensor defined from a (symmetric) metric tensor satisfies $\cR_{\al\be\mu\nu}(g)=-\cR_{\al\be\nu\mu}(g)=-\cR_{\be\al\nu\mu}(g)=\cR_{\mu\nu\al\be}(g)$, the Riemann tensor of a general connection, only has antisymmetry in its two last indices $\cR_{\mu\nu\al\be}(\Ga)\neq\cR_{\al\be\mu\nu}(\Ga)=-\cR_{\al\be\nu\mu}(\Ga)\neq-\cR_{\be\al\nu\mu}(\Ga)$. This implies that, while the Ricci tensor of a (symmetric) metric is always symmetric, the Ricci tensor of a general connection is not. Even if the connection is symmetric, the Ricci tensor will develop an antisymmetric piece.\\

Through the parallel transport operation, any affine connection defines a preferred set of paths named pre-geodesics or auto-parallel paths. Any of these paths can be defined as the equivalence class of curves that are a solution of the parallel transport equation
\beq\label{pregeodesiceq}
\frac{\text{d} x^\mu}{\text{d}\lambda}\na_\mu\lr{\frac{\text{d} x^\nu}{\text{d}\lambda}}=f(\lambda)\frac{\text{d}x^\mu}{\text{d}\lambda},
\eeq
where two curves are equivalent if one is a re-parametrization of the other, or put it in another way, if their image on the manifold is the same set of points. This set of points is the path corresponding to the equivalence class; and the element of the class parametrized in a way such that the right hand side of \eqref{pregeodesiceq} vanishes is the affinely-parametrized geodesic representative of the class.\footnote{Up to an affine transformation $\tau\mapsto \alpha\tau+\beta$. The affine parametrization $\tau(\lambda)$ can always be found by solving $\text{d}^2\tau/\text{d}\lambda^2=(\text{d}\tau/\text{d}\lambda) f(\lambda)$.} Auto-parallel paths are occasionally stipulated to describe the trajectories of test-particles (see however the paragraph below and Section \ref{sec:geodesics} for a detailed discussion on the physical relevance of this stipulation), so it is interesting from the physical point of view to wonder about the existence of a group of transformations acting on the affine connection that is a symmetry of the set of pre-geodesics. Indeed it is known that projective transformations, which act on the connection as\footnote{Notice that the antisymmetric part of the projective transformation $\xi_{[\mu}\delta^\al{}_{\nu]}$  leaves the geodesic equation exactly invariant, whereas the symmetrized part $\xi_{(\mu}\delta^\al{}_{\nu)}$ is the one that leaves the geodesic equation invariant up to a re-definition of the affine parameter.}
\beq\label{projtrans}
\Ga^\al{}_{\mu\nu}\xmapsto{\quad \xi\quad}\bar{\Ga}^\al{}_{\mu\nu}=\Ga^\al{}_{\mu\nu}+\xi_\mu\delta^\al{}_\nu
\eeq
leave pre-geodesics invariant. Here $\xi_\mu$ is any smooth 1-form field. Indeed given a pre-geodesic of $\Gamma^\al{}_{\mu\nu}$, any parametrization of that path is its affine-parametrization for the connection $\bar{\Ga}^\al{}_{\mu\nu}=\Ga^\al{}_{\mu\nu}+A_\mu\delta^\al{}_\nu$  where, as is immediately seen by writing the projectively transformed version of \eqref{pregeodesiceq}, $A_\mu$ has to satisfy $A_\mu\frac{\text{d}x^\mu}{\text{d}\lambda}=-f(\lambda)$ . The meaning of this fact is that, regarding pre-geodesics, a change in the connection by a projective transformation is equivalent to a reparametrization of the solutions of \eqref{pregeodesiceq}, and therefore the equivalence classes of curves related by parametrization (i.e. pre-geodesics) are left invariant under projective transformations. Moreover, at least locally, it is clear from the properties of smooth 1-forms that the projective transformations form a continuous group.

The above discussion implicitly promotes the autoparallel equation to a fundamental level, establishing how particles move in a general affinely connected spacetime. This is not however what occurs in standard physics where the classical equations are derived from an action principle. If we consider a massive test particle, the action is proportional to the line element computed along its trajectory, i.e.,  $\cS=-m\int\dd s$ with $\dd s=\sqrt{\vert g_{\mu\nu}\dot{x}^\mu\dot{x}^\nu\vert}\dd\lambda$. This action is obviously only sensitive to the metric and hence the particle cannot see the affine structure. The equations derived from this action for the particle will be the metric geodesic equations that coincide with the autoparallel equation for the Levi-Civita connection. If we insist in giving a preferred role to the autoparallel equation we should add an explicit coupling to the affine connection in the equations. At a more fundamental level, this amounts to including direct couplings between the connection and the fields. We will come back to these issues in Sec. \ref{sec:geodesics} with a more detailed discussion.

Since the projective symmetry plays a fundamental role in affine geometries and, by extension, in the dynamics of metric-affine theories (as we will see below),  it is useful to unveil the transformation properties of the different tensorial objects associated to the connection under projective transformations. For the sake of generality, let us first write down the transformation law of these connection-related tensorial objects for an arbitrary change in the connection given by
\beq\label{arbitraryconnchange}
\Ga^\al{}_{\mu\nu}\xmapsto{\quad}\Ga^\al{}_{\mu\nu}=\bar{\Ga}^\al{}_{\mu\nu}+\delta\Ga^\al{}_{\mu\nu}.
\eeq
Due to its traditional importance in gravitational theories, let us start by writing down the corresponding transformation rule satisfied by the Riemann curvature tensor and other associated tensors. We find the relation
\begin{align}\label{transfRiemm}
\cR^\al{}_{\be\mu\nu}(\Ga)=\cR^\al{}_{\be\mu\nu}(\bar{\Ga})+2\bar{\na}_{[\mu}\delta\Ga^\al{}_{\nu]\be}+\bar\cT^\lambda{}_{\mu\nu}\delta\Ga^\al{}_{\lambda\beta}+2\delta\Ga^\al{}_{[\mu|\la|}\delta\Ga^\la{}_{\nu]\beta},
\end{align}
where the connection-related objects with an over-bar are defined in terms of the background connection $\bar\Ga^\al{}_{\mu\nu}$. By taking the corresponding traces we are led to
\begin{align}
&\cR_{\mu\nu}(\Ga)=R_{\mu\nu}(\bGa)+2\bar\na_{[\al}\delta\Ga^\al{}_{\nu]\mu}+\bar\cT^\lambda{}_{\al\nu}\delta\Ga^\al{}_{\la\mu}+2\delta\Ga^\al{}_{[\al|\la|}\delta\Ga^\la{}_{\nu]\mu}\;,\label{transfRicci}\\\
&\cP^\mu{}_{\nu}(g,\Ga)=\cP^\mu{}_{\nu}(g,\bGa)+\bar\na_\nu\delta\Ga^{\mu\al}{}_{\al}-\bar\na_\al\delta\Ga^{\mu\;\;\al}_{\;\;\nu}+\bar\cT^\lambda{}_{\nu\al}\delta\Ga^{\mu}{}_{\la}{}^{\al}+2\delta\Ga^\mu{}_{[\nu|\la|}\delta\Ga^{\la}{}_{\al]}{}^{\al}\;,\label{transfcoRicci}\\
&\cQ_{\mu\nu}(\Ga)=\cQ_{\mu\nu}(\bGa)+2\pa{[\mu}\delta\Ga^\al{}_{\nu]\al}\;\label{transfHomothetic}.
\end{align}
Also the changes undergone by the torsion and non-metricity tensors are given by
\begin{align}
&\cT^\al{}_{\mu\nu}(\Ga)=\bar\cT^\al{}_{\mu\nu}+2\delta\Ga^\al{}_{[\mu\nu]}\;,\label{transftor}\\
&Q_{\al\mu\nu}(g,\Ga)=Q_{\al\mu\nu}(g,\bGa)-2\delta\Ga_{(\mu|\al|\nu)}\;,\label{transfnm}
\end{align}
and therefore, the contortion and distortion tensors change as
\begin{align}
&K^\al{}_{\mu\nu}(\Ga)=K^\al{}_{\mu\nu}(\bGa)-\delta\Ga_{(\mu\nu)}{}^\al{}+\delta\Ga^\al{}_{[\mu\nu]}+\delta\Ga_{(\mu}{}^{\al}{}_{\nu)}\;,\label{transfcontor}\\
&L^\al{}_{\mu\nu}(g,\Ga)=L^\al{}_{\mu\nu}(g,\bGa)-\delta\Ga_{(\mu}{}^{\al}{}_{\nu)}+\delta\Ga^\al{}_{(\mu\nu)}+\delta\Ga_{(\mu\nu)}{}^\al\;.\label{transfdistor}
\end{align}
Now that the transformation laws under a general shift in the connection \eqref{arbitraryconnchange} are given, notice that a projective transformation \eqref{projtrans} is a special case of \eqref{arbitraryconnchange} with $\delta\Ga^\al{}_{\mu\nu}=-\xi_\mu\delta^\al{}_{\nu}$. The transformation properties of the different curvature tensors under a projective transformation are then:
\begin{align}\label{ProjetiveTransTensors}
&\cR^\al{}_{\be\mu\nu}(\Ga)=\cR^\al{}_{\be\mu\nu}(\bar{\Ga})-F_{\mu\nu}\delta^\al{}_\be
\quad\quad&\cR_{\mu\nu}(\Ga)=\cR_{\mu\nu}(\bGa)-F_{\mu\nu}\;\nonumber\\
&\cP^\mu{}_{\nu}(\Ga)=\cP^\mu{}_{\nu}(\bGa)+F^{\mu}{}_{\nu}\quad\quad&\cQ_{\mu\nu}(\Ga)=\cQ_{\mu\nu}(\bGa)-DF_{\mu\nu}\;\\
&K^\al{}_{\mu\nu}(\Ga)=K^\al{}_{\mu\nu}(\bGa)+\xi_\nu\delta^\al{}_\mu-g_{\mu\nu}\xi^\al\quad\quad&L^\al{}_{\mu\nu}(g,\Ga)=L^\al{}_{\mu\nu}(g,\bGa)+g_{\mu\nu}\xi^\al-2\xi_{(\mu}\delta^\al{}_{\nu)}\nonumber
\end{align}
where $F_{\mu\nu}=2\pa{[\mu}\xi_{\nu]}$ is the field-strength of the projective mode $\xi$, and $D$ is the number of space-time dimensions. The above transformation laws reveal some interesting properties that will be crucial for the construction of metric-affine theories. Firstly, let us note that projective transformations leave invariant the symmetric parts of the Ricci and co-Ricci tensors, but their antisymmetric part is not. This fact is important because of what follows: It is well known that higher order curvature gravity theories in the metric formalism propagate ghostly degrees of freedom (except Lovelock theories), which can be traced back to the fact that their equations of motion for the metric are of fourth order and present Ostrogradski instabilities. This is due to the fact that, in the metric formalism, if the action has higher order curvature invariants, since the connection contains $\Gamma \supset \pa{}g$, the Riemann and associated curvature tensors contain $R^\alpha{}_{\beta\mu\nu}\supset \pa{}^2g$. Therefore, introducing powers of the Riemann of order higher than one in the action, gives rise to equations of motion for the metric of differential order higher than two. Remarkably, this is not true in the metric-affine formalism, where the connection is no longer related to derivatives of the metric but a fundamental field, and therefore, arbitrary powers of the Riemann in the action do not render higher order equations of motion for the metric. Holding on to this fact, as commented in the Introduction, it is sometimes argued that metric-affine higher order curvature gravity theories do not propagate ghost-like degrees of freedom. This belief is inaccurate because, as explicitly shown in \cite{BeltranJimenez:2019acz}, ghosts arise in theories of gravity whose actions are arbitrary analytic functions of the full Ricci tensor unless further constraints are imposed. We will rederive this results with different complementary approaches below. However, it is also known that metric-affine theories whose actions are an arbitrary analytic function of the symmetric part of the Ricci tensor do not propagate more than the two degrees of freedom of the graviton, and in fact are ghost-free. These two facts can be understood in light of projective symmetry. Indeed, as we have seen above, since $R_{(\mu\nu)}$ is invariant under projective transformations, an action which is an arbitrary function only of the symmetric part of the Ricci enjoys a projective symmetry. On the other hand, an action which is a function of the full Ricci tensor does not because the $R_{[\mu\nu]}$ part explicitly breaks it\footnote{There could be more general actions in which cancellations between different terms contributing to the breaking of projective symmetry lead to a projectively invariant theories, but one would need to consider other objects besides the Ricci tensor.}. We will see that  the breaking of this symmetry unleashes five extra degrees of freedom associated to the presence of the dynamical projective mode with an unavoidable ghostly sector. This poses a serious drawback to consider general non-projectively invariant gravity theories. There is a loophole in the argument which we will explore: By introducing additional constraints, even theories which break projective symmetry can be safe. Let us now introduce the general formulation of RBGs.

%%%%%%%%%%%%%%%%%%%%%%%%%%%%%%%%%%%%%%%%%%%%%%%%%%%%%%%%%%%%%%%%%%%%%%%

\section{Ricci-based metric-affine theories}\label{sec:RBG}
Equipped with the general geometrical framework introduced in the previous section, we can turn to the family of theories that will conform the main focus of this work, namely theories of gravity formulated in the metric-affine approach and that only depend on the Ricci tensor. This might give the impression of an unnecessary restraint given the huge freedom permitted by the the general metric-affine formalism. Let us recall at this point that we have a plethora of different geometrical objects that could be used and which should indeed enter the action, unless some additional guiding principle is invoked. Since our purpose here is showing the (generically) pathological nature of higher order curvature theories of gravity in the metric-affine formalism, we simply take these theories as a benchmark to illustrate the potential problems suffered by metric-affine theories. It is important however to stress that RBG theories have received considerable attention in the literature \cite{Afonso:2017bxr,BeltranJimenez:2017doy,Afonso:2018bpv,Afonso:2018hyj,Afonso:2018mxn,Delhom:2019zrb,Latorre:2017uve,Delhom:2019wir}, due to their interesting properties that make them appealing and more tractable than other more general metric-affine theories, thus being useful as a proxy to better understanding general metric-affine theories.

\subsection{Field equations}\label{sec:FieldEqsRBGgeneral}
The family of theories that we will mainly consider throughout our analysis will be described by an action of the following form:
\beq\label{GeneralAction}
\cS[g_{\mu\nu},\Ga^\al{}_{\mu\nu},\Psi]=\frac12\int^{}_{} \dd^Dx \sqrt{-g}\,F\big(g^{\mu\nu},\cR_{\mu\nu}\big)+\cS_{\rm m}[g_{\mu\nu},\Psi],
\eeq
where $F$ is an arbitrary scalar function that depends on the (inverse) metric $g^{\mu\nu}$ and the Ricci tensor $\cR_{\mu\nu}$ of an arbitrary connection $\Gamma^\alpha{}_{\mu\nu}$ that is to be determined by the field equations. We have also included the matter sector through its action $\cS_{\rm m}$, where $\Psi$ stands for all matter fields, i.e., it can carry both internal and Lorentz indices. Unless otherwise stated, we will  assume that matter fields are minimally coupled to gravity. In the metric formalism, this is a pretty straightforward procedure to follow free from ambiguities. However, in the metric affine-formalism, even this prescription leads to ambiguities in several respects that lead to the appearance of terms involving the torsion and/or non-metricity tensors (see e.g. \cite{Delhom:2020hkb,BeltranJimenez:2020sih}). For the moment, we will have in mind matter actions containing scalar fields with up to first derivatives and vector fields whose kinetic terms are gauge invariant. That way the matter sector will not contain the connection and all the dependence on $\Gamma^\alpha{}_{\mu\nu}$ will come from the gravitational sector. We will drop this assumptions later and discuss their impact.

The field equations obtained by varying \eqref{GeneralAction} with respect to the metric and the connection are respectively
\begin{eqnarray}
\frac{\partial F}{\partial g^{\mu\nu}}-\frac{1}{2}F g_{\mu\nu}&=&T_{\mu\nu}\label{MetricFieldEqs}\\
\na_\la\lrsq{\sqrt{-q}q^{\nu\mu}}-\del^\mu{}_\la\na_\rho\lrsq{\sqrt{-q}q^{\nu\rho}}
&=&\sqrt{-q}\lrsq{\cT^\mu{}_{\la\al} q^{\nu\al}+\cT^\al{}_{\al\la} q^{\nu\mu}-\del^\mu{}_\la\cT^\al{}_{\al\be} q^{\nu\be}},\label{VariationConnection}
\end{eqnarray}
where $T_{\mu\nu}=-\frac{2}{\sqrt{-g}}\frac{\delta\cS_{\rm m}}{\delta g^{\mu\nu}}$ is the usual energy-momentum tensor of the matter sector and we have introduced the object $\sqrt{-q}q^{\mu\nu}\equiv\sqrt{-g}\frac{\partial F}{\partial \cR_{\mu\nu}}$. In the usual treatment of RBGs, projective symmetry is assumed, which from \eqref{ProjetiveTransTensors} restricts the dependence of the action only to the symmetric part of the Ricci tensor. Here, we want to offer a more detailed discussion about what are the consequences of breaking the projective symmetry in RBGs than that presented in \cite{BeltranJimenez:2019acz}. Thus in our discussion  the full Ricci tensor will enter the action and, therefore, the object $q^{\mu\nu}$ will carry all its 16 components instead of the 10 components of the projectively symmetric case\footnote{Note that in the projectively symmetric case, as only the symmetric part of the Ricci tensor enters the action, the corresponding auxiliary metric $q^{\mu\nu}$ is a symmetric tensor.}. Turning back to the field equations of the generalised RBGs, the above connection equation can be recast in a more useful form by introducing a new connection $\hat{\Gamma}^\alpha{}_{\mu\nu}$ obtained by subtracting a projective mode from the original one
\beq
\hat{\Gamma}^\alpha{}_{\mu\nu}=\Gamma^\alpha{}_{\mu\nu}+\frac{2}{D-1}\Gamma^\lambda{}_{[\lambda\mu]}\delta^\alpha{}_\nu,
\label{Eq:Gammaheq}
\eeq
which identically satisfies $\hat{\Gamma}^\lambda{}_{[\lambda\mu]}=0$. In terms of this new connection, \eqref{VariationConnection} can be recast as
\beq
\partial_\la(\sqrt{-q}q^{\mu\nu})+\hat{\Gamma}^{\mu}{}_{\la\al}\sqrt{-q}q^{\al\nu}+\hat{\Gamma}^{\nu}{}_{\al\la}\sqrt{-q}q^{\mu\al}-\hat{\Gamma}^{\al}{}_{\la\al}\sqrt{-q}q^{\mu\nu}=0.\label{conneq}
\eeq
We can now remove the different traces of the connection appearing in the field equations by taking the different traces of the above equation. By doing so, we get from the algebraic manipulations the condition
\beq
\partial_\mu\Big(\sqrt{-q} q^{[\mu\nu]}\Big)=0,
\eeq
and then we arrive at the same connection equation found in \cite{Damour:1991ru} for NGT:
\beq
\partial_\lambda q^{\mu\nu}+\hat{\Gamma}^\mu{}_{\lambda\al}q^{\al\nu}+\hat{\Gamma}^\nu{}_{\al\lambda}q^{\mu\al}=0.
\label{Eq:Gammah}
\eeq

Solving Eq. (\ref{Eq:Gammaheq}) for the connection is in general quite cumbersome, if possible at all. What is easy to see is that this equation can be algebraically solved in terms of $q^{\mu\nu}$. However, $q^{\mu\nu}$ as defined above depends itself on the connection through its dependence on $\cR_{\mu\nu}$ so that this does not in general give the solution for the connection. A singular case is when the function is linear in the Ricci tensor so that $q^{\mu\nu}$ does not depend on the connection. This is of course the case for the Einstein-Hilbert action. Thus, although it would be possible to work directly with the above equations, and indeed we will solve them perturbatively in the antisymmetric part of $q^{\mu\nu}$, it is useful to consider other ways of working with generalised RBGs. Indeed it can be seen that all RBG theories can be described in terms of an Einstein-Hilbert-like term $q^{\mu\nu}\cR_{\mu\nu}$ by performing a suitable field re-definition. Let us clarify this in the next section.

%%%%%%%%%%%%%%%%%%%%%%%%%%%%%%%%%%%%
\subsection{The Einstein-Hilbert frame}\label{sec:NSGFrame}

As discussed above, it is possible to obtain the main properties of general RBG theories by working with the field equations. However, it is more illuminating to re-write the action so that the gravitational sector looks more familiar and, consequently, the physical content of the theory is more apparent. We will follow the procedure presented in \cite{BeltranJimenez:2017doy,Afonso:2017bxr} for the projectively invariant theories, extending it to the general non-projectively invariant case. Let us start by performing a Legendre transformation in order to linearise the action in the Ricci tensor as follows:
\beq\label{ModifiedAction}
\cS=\frac12\int \dd^Dx \sqrt{-g}\left[F(\Sigma_{\mu\nu})+\frac{\partial F}{\partial \Sigma_{\mu\nu}}\big(\cR_{\mu\nu}-\Sigma_{\mu\nu}\big)\right]
+\cS_{\rm m}[g,\Psi],
\eeq
where $\Sigma_{\mu\nu}$ is an auxiliary field. Unlike in the projectively invariant case, where $\Sigma_{\mu\nu}$ is symmetric, this auxiliary field does not have any symmetry carrying both symmetric and antisymmetric parts. We will see below that it is precisely the antisymmetric part that gives rise to the pathological behaviour of these theories. In order to put our action in a more familiar form, we will introduce the following field re-definition: 
\beq
\sqrt{-q}q^{\mu\nu}=\sqrt{-g}\frac{\partial F}{\partial \Sigma_{\mu\nu}}.
\eeq
This definition will allow to express the auxiliary field $\Sigma_{\mu\nu}$ in terms of the spacetime metric and the object $q^{\mu\nu}$, i.e., we will have an algebraic relation $\Sigma_{\mu\nu}=\Sigma_{\mu\nu}(q,g)$. The resemblance of this definition with the one introduced in the field equations of sec. \ref{sec:FieldEqsRBGgeneral} is due to the fact that the dynamics of this new auxiliary field is given by the constraint $\Sigma_{\mu\nu}=\cR_{\mu\nu}$, so that the above field re-definition looks exactly like the definition for $q^{\mu\nu}$ given in the previous section when the field equations are satisfied. After this field re-definition, we can then express the RBG action in the form
\beq\label{NonSymmetricActionTwoMetrics}
\cS=\frac12\int \dd^Dx\Big[\sqrt{-q}q^{\mu\nu}\cR_{\mu\nu}(\Ga)+\cU(q,g)\Big]+\cS_{\rm m}[g,\Psi],
\eeq
where we have introduced the potential term
\beq
\cU(q,g)=\sqrt{-g}\lrsq{F-\frac{\partial F}{\partial\Sigma_{\mu\nu}}\Sigma_{\mu\nu}}_{\Sigma=\Sigma(q,g)}.
\eeq
The action \eqref{NonSymmetricActionTwoMetrics} already features the standard Einstein-Hilbert term in the first order formalism, but for the object $q^{\mu\nu}$ instead of the spacetime metric $g_{\mu\nu}$. As a matter of fact, we can notice that $g_{\mu\nu}$ appears algebraically in the potential $\cU$ and the matter action so that it is simply an auxiliary field that we can integrate out by solving its equation of motion
\beq
\frac{\partial \cU}{\partial g^{\mu\nu}}=\sqrt{-g}\,T_{\mu\nu}.
\label{Eq:metric}
\eeq
From this equation we can obtain the spacetime metric $g_{\mu\nu}$ in terms of the object $q^{\mu\nu}$ and the energy-momentum tensor of the matter sector, computed as the variation of the matter action w.r.t. $g_{\mu\nu}$ as usual. We will see below that there is another energy-momentum tensor that we can introduce to make the resemblance with the first-order formulation of GR even more apparent. Once we have obtained the corresponding solution to \eqref{Eq:metric}, we can use it to finally express \eqref{NonSymmetricActionTwoMetrics} as
\beq
\cS=\frac12\int \dd^Dx \Big[\sqrt{-q}q^{\mu\nu}\cR_{\mu\nu}(\Ga)+\cU(q,T)\Big]+\cS_{\rm m}[g(q,T),\Psi].
\label{Eq:EHframe1}
\eeq
This is the desired appearance of the theory where the gravitational sector reduces to the well-known Einstein-Hilbert action in the first order formalism. It is important to emphasise that the resemblance is purely formal at this point and, in fact, solving for the connection will fail to recover GR owed to the lack of any symmetries of $q^{\mu\nu}$. In the next sections we will explicitly show when this is the case and the differences when it is not. 

In addition to the purely gravitational sector, we also see how we have generated couplings between the object $q^{\mu\nu}$ and the matter sector. Such couplings arise from two sources after integrating out the spacetime metric, namely: from the potential $\cU$ generated when linearising in the Ricci tensor and from the explicit couplings of the matter sector to $g_{\mu\nu}$. Notice that, since the matter sector was assumed to be minimally coupled to gravity, i.e., it only couples to $g_{\mu\nu}$, matter will only enter Eqs. (\refeq{Eq:metric}) through the energy-momentum tensor obtained as the reaction to variations of the spacetime metric. This further implies that all the newly generated matter couplings will only depend on $T_{\mu\nu}$, which guarantees the preservation of the symmetries in the original matter sector. Notice that since $g_{\mu\nu}$ appears in $T_{\mu\nu}$ not as $T_{\mu\nu}\propto g_{\mu\nu}$ but in a more involved form, it could be that if we truly want to eliminate $g_{\mu\nu}$ in favour of $q_{\mu\nu}$ and the matter fields, the dependence could also be more general than through $T_{\mu\nu}$ (we have to solve the corresponding equation for $g_{\mu\nu}$). However, the new couplings will still surelly have the same symmetries as the matter action.

%%%%%%%%%%%%%%%%%%%%%%%%%%%%%%%%%%%%%%%%%%%%%%%%%%%%%%%%%%%%
%%%%%%%%%%%%%%%%%%%%%%%%%%%%%%%%%%%%%%%%%%%%%%%%%%%%%%%%%%%%
%%%%%%%%%%%%%%%%%%%%%%%%%%%%%%%%%%%%%%%%%%%%%%%%%%%%%%%%%%%%%%%%%%%%%%%%
\section{Projectively-invariant theories: Equivalence to GR}\label{sec:PItheo}

Before proceeding to the general case where the object $q^{\mu\nu}$ does not exhibit any symmetries, let us consider what happens when a projective symmetry is imposed. This has already been studied in the literature, but it will be useful to discuss these known results here in order to appreciate better the fundamental role played by the projective mode in these theories. As explained in Sec. \ref{sec:Generalities}, in Ricci-Based actions, the projective symmetry can be straightforwardly implemented by restricting the action to depend only on the symmetric part of the Ricci tensor. If that is the case, then it is easy to see from the definition of $q^{\mu\nu}$ that this object will inherit the symmetric character of the Ricci tensor. Being a symmetric rank-2 tensor, $q^{\mu\nu}$ is then entitled to claim its status as a proper metric tensor so that the gravitational sector in (\ref{Eq:EHframe1}) is actually the first order formulation of GR. However, the corresponding solution for the connection will be given by the Christoffel symbols of the metric $q^{\mu\nu}$ (up to the projective mode entering as a gauge mode \cite{Bernal:2016lhq}) instead of those of the spacetime metric $g_{\mu\nu}$.

In the Einstein-frame we thus recover the usual form of the Einstein equations, but the right hand side is now given by the energy-momentum tensor describing the reaction to the metric $q^{\mu\nu}$ of the matter action resulting after integrating out $g_{\mu\nu}$, i.e.
\beq
\tilde{T}_{\mu\nu}=-\frac{2}{\sqrt{-q}}\frac{\delta \tilde{S}_{\rm m}}{\delta q^{\mu\nu}}.
\eeq
This energy-momentum is highly non-linearly related to $T_{\mu\nu}$ \cite{Afonso:2018bpv}, and will feature new interactions between all the matter fields in general \cite{Latorre:2017uve,Delhom:2019wir}, which are the origin of the different phenomenology and solutions that differ from the {\it usual GR} behaviour. Let us stress however that these theories are nothing but standard GR in disguise. The apparent differences between RBGs and GR are simply due to the fact that a matter sector coupled to a projectively invariant RBG corresponds to another matter sector (obtained as a non-linear deformation of the previous one) coupled to GR. The peculiar property of the RBG with projective symmetry is that the interactions in the matter sector present a somewhat universal form (that of course depends on the specific theory, i.e., the function $F$). As we have discussed above, if we start from minimally coupled matter fields, all the new interactions will be generated through the total energy-momentum tensor \cite{Afonso:2018bpv,Delhom:2019wir}. Assuming that the most relevant interactions in the gravitational sector of RBG appear at some specific scale $\Lambda$, which means that the function $F$ only contains one additional dimensionfull parameter, then all the new interactions in the matter sector will not only be universally constructed in terms of $T_{\mu\nu}$, but they all will in turn have the same coupling constant. This means that, if an effect is seen at a given scale in some sector of the standard model, effects at the same scale will arise in the remaining sectors. Regarded from this perspective in the Einstein frame, we can interpret RBG theories as a procedure to encapsulate a universally interacting matter sector, in the sense explained above, in an auxiliary field that plays the role of a non-dynamical connection. In particular, this property is precisely what permits to study the dynamics in terms of a metric $g_{\mu\nu}$ for all matter fields at the same time. Let us elaborate on this point a bit more.

The physical meaning of the two metrics is also apparent in the Einstein frame, again assuming minimally coupled fields. The metric $g_{\mu\nu}$ will determine the trajectories of the particles, which will follow the corresponding geodesics\footnote{It is perhaps convenient to explicitly state the physical situation we have in mind and what we mean by particles and geodesics. We assume that there is some background configuration both for the gravitational sector and the matter fields. Then, there will be perturbations on top of this background configuration and these perturbations are what we will call {\it particles}, possibly with an unfortunate abuse of language. These perturbations are the ones that will follow geodesics of a given metric when we consider their {\it free} propagation. Of course, living on a non-trivial background, the propagation will occur in a medium with which these perturbations will interact.}. One may then wonder why they do not follow the geodesics of $q_{\mu\nu}$ in the Einstein-frame and how to square this with our statement that these theories are GR. The answer is quite simple. Around trivial matter backgrounds, both metrics are the same and therefore there is no possible confusion. In the presence of a matter background however both metrics are different and while matter fields follow the geodesics of $g_{\mu\nu}$, it is $q_{\mu\nu}$ that satisfies Einstein equations. There is no onus however because, also in GR when matter fields propagate on a non-trivial background (and are coupled to it) the propagation does not follow the geodesics of $g_{\mu\nu}$. Paradigmatic examples of this behaviour are for instance $K$-essence models of scalar fields or non-linear electrodynamics (see e.g. \cite{Babichev:2007dw,ArmendarizPicon:2005nz,Plebanski:106680,Novello:1999pg,Gibbons:2000xe}). As a matter of fact, starting from a standard canonical scalar field and usual Maxwellian electrodynamics in the RBG frame, the Einstein frame formulation will precisely be $K$-essence \cite{Afonso:2018hyj} and non-linear electrodynamics respectively \cite{Delhom:2019zrb}.

%%%%%%%%%%%%%%%%%%%%%%%%%%%%%%%%%%%%%%%%%%%%%%%%%%%%%%%%%%%%
%%%%%%%%%%%%%%%%%%%%%%%%%%%%%%%%%%%%%%%%%%%%%%%%%%%%%%%%%%%%
\section{Generalised RBG theories: The Non-symmetric Gravity frame}\label{sec:GenRBG}
The explicit breaking of projective symmetry in the RBG Lagrangian allows the full Ricci tensor to appear in the action, thus jeopardising the symmetric nature of the corresponding $q^{\mu\nu}$. This crucially changes the situation  and the resulting theory in the Einstein frame representation is no longer GR but it resembles the Nonsymmetric Gravity Theory  (NGT) introduced by Moffat \cite{Moffat:1978tr} and which has been explored in different versions. Although the non-symmetric frame of generalised RBGs does not exactly reproduce Moffat's non-symmetric gravity, it does so in certain limits. A crucial difference is the coupling to matter fields, although even this can be made equivalent by ad-hoc choices of the matter couplings in Moffat's theory. Thus, given the similarities between both theories, it will be instructive to review some of the known results on non-symmetric gravity that can then be straightforwardly applied to the generalised RBGs. In particular, we will review the pathologies that plague Moffat's theory \cite{Damour:1991ru,Damour:1992bt} (see also \cite{Moffat:1993gi,Moffat:1994hv,Clayton:1995yi,Clayton:1996dz,Moffat:1996hf,Prokopec:2005fb,Poplawski:2006ev,Janssen:2006nn,Janssen:2006jx,BeltranJimenez:2012sz,Golovnev:2014aca}) and how they will then be inherited by generalised RBGs. We will seize the opportunity to provide alternative understandings for the origin of the pathologies. Let us start by considering vacuum solutions so that no matter fields are present\footnote{We allow the appearance of a cosmological constant like term $\bar\cU$ that accounts for a possible non-trivial dependence of $\cU$ on the background $q^{\mu\nu}$ solution.} and the analysis of the gravitational sector becomes cleaner. Thus, the starting action for NGT (or generalised RBGs in the Einstein-Hilbert frame) will be
\beq
\cS=\frac12\int \dd^Dx \Big[\sqrt{-q}\mpl^2q^{\mu\nu}\cR_{\mu\nu}(\Ga)+\bar\cU\Big],
\eeq
where $q^{\mu\nu}$ is a metric with an antisymmetric part\footnote{Notice the formal equivalence with \eqref{Eq:EHframe1} when the projective symmetry is broken and the corresponding $q^{\mu\nu}$ develops an antisymmetric part.} and $\cU$ is some potential for the non-symmetric object $q_{\mu\nu}$. Of course, in the case of a symmetric $q_{\mu\nu}$, this term can only contribute a cosmological constant by virtue of covariance, but it can have a non-trivial structure for the non-symmetric case with important consequences. In fact, such a term was invoked in \cite{Damour:1992bt} to resolve the pathologies of Moffat's theory. The instabilities that plague this theory around arbitrary backgrounds can be evidenced by different methods that provide complementary views. Let us start by the allegedly simplest procedure to show the presence of pathologies.

\subsection{Instabilities in the decoupling limit}\label{sec:InstDecLim}
We will first study a suitable decoupling limit of the theory that already manifests the presence of ghosts. For that, we will consider the antisymmetric sector perturbatively up to quadratic order so that
\beq
q_{\mu\nu}=\sq_{\mu\nu}+\frac{\sqrt{2}}{\mpl} \left(B_{\mu\nu}+\alpha B_{\mu\alpha} B^\alpha{}_\nu+\beta B^2\sq_{\mu\nu}\right),
\eeq
with $\sq_{\mu\nu}$ an arbitrary symmetric metric, $B_{\mu\nu}$ a 2-form field corresponding to the antisymmetric part of $q_{\mu\nu}$, and where the parameters $\alpha$ and $\beta$ account for the possibility of field re-definitions at quadratic order (see e.g. \cite{Damour:1992bt}). The numerical factor and the Planck mass have been introduced for convenience. When expanding around such a background at second order in $B_{\mu\nu}$ we find\footnote{Here we will stick to the $D=4$ case for simplicity. In arbitrary dimensions, the analysis can be carried in a similar fashion, although taking into account that the degrees of freedom carried by each field might change with the dimension.}:
\begin{align}
\cS^{(2)}=\int\dd^4x\sqrt{-\sq}\Big[&\frac12\mpl^2R(\sq)-\frac{1}{12} H_{\mu\nu\rho} H^{\mu\nu\rho}-\frac14 m^2 B^2-\frac{\sqrt{2}\mpl}{3}B^{\mu\nu}\partial_{[\mu}\Gamma_{\nu]}\nonumber\\
&+\frac{1-2\alpha+4\beta}{4} R(\sq) B^2+\alpha R_{\mu\nu}(\sq)B^{\mu\alpha}B^\nu{}_\alpha-R_{\mu\nu\alpha\beta}(\sq)B^{\mu\alpha}B^{\nu\beta}\Big]
\label{eq:Actquad}
\end{align}
where $H_{\mu\nu\rho}=3\partial_{[\mu} B_{\nu\rho]}$ the field strength of the 2-form field, $m^2$ is the mass generated from $\bar\cU$, and $\Gamma_\mu$ is the projective mode of the connection. In order to make apparent the presence and nature of the instabilities, we will first follow a different approach from those used in analysis of NGT that will allow us to clearly pinpoint the problems, namely we will resort to the St\"uckelberg trick. Let us first consider a flat background so the couplings to curvature in (\ref{eq:Actquad}) disappear. Then, we can restore the gauge symmetry of the 2-form by introducing St\"uckelberg fields $b_\mu$ via the replacement $B_{\mu\nu}\rightarrow \hat{B}_{\mu\nu}+\frac2m \partial_{[\mu} b_{\nu]}$, and take the decoupling limit $m\rightarrow 0$. There will still be the scalar mode of the gauge invariant 2-form sector described by $\hat{B}_{\mu\nu}$ that we do not need to consider to show the presence of a ghost. The relevant sector in the decoupling limit of the action in a flat background is then
\begin{align}\label{ActionDec}
\cS^{(2)}_{\rm dec, flat}=-\int\dd^4x\sqrt{-\sq}\left(\frac{1}{12} \hat{H}_{\mu\nu\rho} \hat{H}^{\mu\nu\rho}+\frac14 \cB_{\mu\nu} \cB^{\mu\nu}+ \cB_{\mu\nu} \Gamma^{\mu\nu} \right)
\end{align}
where $\hat{H}_{\mu\nu\rho}=3\partial_{[\mu} \hat{B}_{\nu\rho]}$, $\cB_{\mu\nu}=2 \partial_{[\mu} b_{\nu]}$ and $\Gamma_{\mu\nu} =2 \partial_{[\mu} \Gamma_{\nu]}$. In order to properly take the decoupling limit, we have re-scaled $\Gamma_\mu\rightarrow \frac{3m}{\sqrt{2}\mpl}\Gamma_\mu$ that has been kept finite. We see that the decoupling limit shows the presence of five degrees of freedom, namely: one associated to the massless 2-form $\hat{B}_{\mu\nu}$ and two associated to the helicity-1 modes described by $b_\mu$ and the projective mode. This is of course the expected counting for \eqref{eq:Actquad} corresponding to a massive 2-form and a gauge spin-1 field. In this decoupling limit it is then apparent that the theory is plagued by ghost-like instabilities owed to the mixing  $\cB_{\mu\nu} \Gamma^{\mu\nu}$ that comes in without the diagonal $ \Gamma_{\mu\nu} \Gamma^{\mu\nu}$ element. This signals the presence of a ghost caused by the negative definite character of the kinetic matrix. More explicitly, if we diagonalise by means of $b_\mu=A_\mu+\xi_\mu$, $\Gamma_\mu=\lambda A_\mu-(2+\lambda) \xi_ \mu$,  the action \eqref{ActionDec} reads
\beq
\cS^{(2)}_{\rm dec,flat}=-\int\dd^4x\sqrt{-\sq}\left[\frac{1}{12} \hat{H}_{\mu\nu\rho} \hat{H}^{\mu\nu\rho}+ \frac{1+\lambda}{4}\Big(\partial_{[\mu} A_{\nu]} \partial^{[\mu} A^{\nu]}- \partial_{[\mu} \xi_{\nu]} \partial^{[\mu} \xi^{\nu]}\Big)\right],
\eeq
 showing that either $A_\mu$ or $\xi_\mu$ is necessarily a ghost. We have reproduced the result announced in \cite{BeltranJimenez:2019acz} in the decoupling limit of the theory.

After showing the presence of a ghost in a flat background,  we will turn on the symmetric sector and allow for an arbitrary curved $\bar{q}$-background. It should then be clear that the non-minimal couplings to the curvature in (\ref{eq:Actquad}) will present additional pathologies. These pathologies have also been discussed for NGT in \cite{Damour:1992bt}. Within our approach we can readily see and interpret the nature of these pathologies as Ostrogradski instabilities \cite{Woodard:2015zca} associated to having higher order equations of motion for the Stueckelberg fields\footnote{The Ostrogradski instabilities have not been properly identified within NGT and represent yet another problem for NGT besides the pathological asymptotic fall off behaviour discussed in \cite{Damour:1992bt}.}. The appropriate decoupling limit now needs to take into account that the curvature scales as $R\sim\mpl^{-2}$ and the appropriate limit to be taken is $m\rightarrow0$ and $\mpl\rightarrow\infty$ with $\Lambda\equiv m \mpl$ fixed. In this limit, the St\"uckelberg fields $b_\mu$ will feature non-minimal couplings with the schematic form $\sim \frac{1}{\Lambda^2}R\cB\cB$. It is known that these derivative couplings generically give rise to higher order equations of motion, thus giving rise to Ostrogradski instabilities. An exceptional case is provided by the Horndeski vector-tensor interaction found in \cite{Horndeski:1976gi}. Having the two free parameters $\alpha$ and $\beta$ that allow for field redefinitions at quadratic order, one would be tempted to say that the pathology is not physical since the Horndeski interaction could be reached by an appropriate local field redefinition. It is worth noticing that even this Horndeski interaction presents pathologies around relevant backgrounds \cite{Jimenez:2013qsa}.  Nevertheless, we need to remember that this is the quadratic action and it is expected that going to higher perturbative orders, new higher order non-minimal couplings will be generated. Since there are no healthy such terms beyond the Horndeski interaction in four dimensions, these will need to be trivial modulo field redefinition to avoid re-introducing the pathologies. At this point, the pathological character of these theories should be unequivocal taken at face value. One could argue that interpreted as effective field theories, there could be a certain regime of validity at low energies. However, the very presence of the ghosts already around a Minkowski background shown above makes this hope difficult to realise. In this respect, this ghost could be stabilised easily by introducing a term $\Gamma_{\mu\nu} \Gamma^{\mu\nu}$. Although such a term cannot be generated from RBGs, within an EFT approach, not only it should appear, but also a bunch of other terms accompanying it.\footnote{An EFT approach to the restricted class of Poincar\'e gauge theories has been pursued in \cite{Aoki:2019snr}.} The non-minimal couplings however, being (irrelevant) higher dimension operators, should typically be perturbative and, consequently, the associated ghosts would only come at a scale beyond the cut-off. One could tune some coefficients to push the ghosts to higher scales so that the corresponding irrelevant operators could have non-perturbative effects on the low-energy phenomenology. This is clearly beyond the scope of this works, but it would be an interesting study to pursue.  

A potential caveat of our analysis (up to now) is that we have neglected the matter sector, but this should not worry us too much since including matter fields will hardly render the theories stable. Rather, one could expect a more pathological behaviour. We will address this point later to show it explicitly. 

To summarise, we have seen that the breaking of the projective symmetry results in the appearance of five degrees of freedom, two of which correspond to the projective mode and the remaining three belong to the antisymmetric part of the metric. In both sectors we have clearly identified the root for the problems and we can now understand that it is precisely the trivialisation of the affinity in projectively-invariant RBG theories what makes them viable by reducing their gravitational sector to GR.

%%%%%%%%%%%%%%%%%%%%%%%%%%%%%%%%%%%%%%%%%%%%%%%%%%%%%%%%%%%%
%%%%%%%%%%%%%%%%%%%%%%%%%%%%%%%%%%%%%%%%%%%%%%%%%%%%%%%%%%%%
%%%%%%%%%%%%%%%%%%%%%%%%%%%%%%%%%%%%%%%%%%%%%%%%%%%%%%%%%%%%
%%%%%%%%%%%%%%%%%%%%%%%%%%%%%%%%%%%%%%%%%%%%%%%%%%%%%%%%%%%%

\subsection{Another view on the problem with additional dofs}\label{sec:ProblemAdditionalDOFs}
In the previous section we have shown how vacuum RBG without a projective symmetry (or vacuum NGT for that matter) are plagued by ghost-like instabilities arising from two sectors, namely: the dynamical projective mode whose mixing with the 2-form leads to the necessary presence of a spin-1 ghost and the non-minimal couplings of the 2-form field that gives rise to Ostrogradski instabilities. This has been neatly shown in the decoupling limit of the theories. Here we will show the appearance of these pathologies in an alternative manner. Let us consider our family of theories described by the action
\beq\label{GeneralAction2}
\cS[g_{\mu\nu},\Ga]=\frac12\int \dd^Dx \sqrt{-g}\,F\big(g^{\mu\nu},\cR_{\mu\nu}(\Ga)\big),
\eeq
where we again consider vacuum generalised RBGs. Let us now separate a metric contribution to the connection from the rest, i.e., let us perform the following field re-definition 
\beq
\Gamma^\alpha{}_{\mu\beta}=\Ch{\alpha}{\mu}{\beta}(h)+\Upsilon^\alpha{}_{\mu\beta}
\label{Eq:GammasplittingChO}
\eeq
where $\Ch{\alpha}{\mu}{\beta}(h)$ are the Christoffel symbols of a metric that we have called $h_{\mu\nu}$ and that we will choose in a convenient manner. After splitting the non-symmetric metric as
\beq\label{metricsplitting}
\sqrt{-q}q^{\mu\nu}=\sqrt{-h}h^{\mu\nu}+\sqrt{-h}B^{\mu\nu}
\eeq
with $\sqrt{-h}h^{\mu\nu}=\sqrt{-q}q^{(\mu\nu)}$ and $\sqrt{-h}B^{\mu\nu}=\sqrt{-q}q^{[\mu\nu]}$, and using (\ref{transfRicci}) for the field re-definition (\ref{Eq:GammasplittingChO}), we can write the generalised RBG action in its Einstein-Hilbert frame \eqref{Eq:EHframe1} as
\beq
\begin{split}
\cS=\frac12\int\dd^Dx\sqrt{-h}\Big[&R(h)-\Upsilon^{\lambda\alpha\mu}\Upsilon_{\alpha\mu\lambda}+\Upsilon^\alpha{}_{\alpha\lambda}\Upsilon^\lambda{}_{\kappa}{}^\kappa-\Upsilon^\alpha{}_{\alpha\lambda}\Upsilon^\lambda{}_{\mu\nu}B^{\mu\nu}\\&
-\Upsilon^\alpha{}_{\nu\lambda}\Upsilon^\lambda{}_{\alpha\mu}B^{\mu\nu}-B^{\mu\nu}\nabla^h_\alpha\Upsilon^\alpha{}_{\mu\nu}-B^{\mu\nu}\nabla^h_\nu\Upsilon^\alpha{}_{\alpha\mu}+\cU(B)\Big].
\end{split}
\eeq
Here we have used the fact that the connection $\Ch{\alpha}{\mu}{\beta}$ is torsion-free, $\na^h$ is the covariant derivative with respect to the Levi-Civita connection of $h^{\mu\nu}$, and we have dropped a boundary term. Notice that we have used (and will use in the subsequent manipulations) $h_{\mu\nu}$ as the metric so we will raise and lower indices with $h^{\mu\nu}$ and its inverse $h_{\mu\nu}$. The field equations for $\Upsilon^\alpha{}_{\mu\nu}$ obtained by variation of the above action are
\beq
B^{\mu\nu} \Upsilon^{\beta}{}_{\beta\alpha}  - h^{\mu\nu} \Upsilon^{\beta}{}_{\beta\alpha} + 
B^{\nu}{}_{\beta} \Upsilon^{\mu\beta}{}_{\alpha} + \Upsilon^{\mu\nu}{}_{\alpha} - B^{\mu}{}_{\beta} 
\Upsilon^{\nu}{}_{\alpha}{}^{\beta} + \Upsilon^{\nu}{}_{\alpha}{}^{\mu}  - 
\delta_{\alpha}{}^{\mu} \Upsilon^{\nu\beta}{}_{\beta} + B_{\beta\lambda} \delta_{\alpha}{}^{\mu} 
\Upsilon^{\nu\beta\lambda}  - \nabla^h_{\alpha}B^{\mu\nu}  - \delta_{\alpha}{}^{\mu} \nabla^h_{\beta}B^{\nu\beta}=0,
\label{Eq:Omegaeq}
\eeq
and taking the trace with respect to $\alpha$ and $\nu$ of this equation we obtain
\beq\label{DivConstB}
\nabla^h_\mu B^{\mu\nu}=0
\eeq
which constrains the 2-form field $B^{\mu\nu}$ to be divergence-free and leaves the connection equation as
\beq
\nabla^h_{\alpha}B^{\mu\nu} - B^{\mu\nu} \Upsilon^{\beta}{}_{\beta\alpha}  -
B^{\nu}{}_{\beta} \Upsilon^{\mu\beta}{}_{\alpha} + B^{\mu}{}_{\beta} 
\Upsilon^{\nu}{}_{\alpha}{}^{\beta} - B_{\beta\lambda} \delta_{\alpha}{}^{\mu} 
\Upsilon^{\nu\beta\lambda} + h^{\mu\nu} \Upsilon^{\beta}{}_{\beta\alpha} - \Upsilon^{\mu\nu}{}_{\alpha} - \Upsilon^{\nu}{}_{\alpha}{}^{\mu}  + 
\delta_{\alpha}{}^{\mu} \Upsilon^{\nu\beta}{}_{\beta}   =0,
\label{Eq:Omegaeqdivfree}
\eeq
This constraint on the 2-form shows that (in $D=4$) $B^{\mu\nu}$ can be expressed as the dual of the field strength of some 1-form $A_\mu$ so that we can write $B^{\mu\nu}=-\frac{1}{2\sqrt{-h}}\epsilon^{\mu\nu\alpha\beta}\partial_{[\alpha}A_{\beta]}$. Notice that the constraint is exact so that we see that the 2-form can propagate at most the same number of degrees of freedom as a vector field (see Appendix \ref{sec:Appendix2-form}). It is also easy to see that a projective mode $\Upsilon^\alpha{}_{\mu\nu}=\xi_\mu\delta^\alpha{}_\nu$ is a solution when $B_{\mu\nu}=0$. This was indeed expected since for vanishing $B_{\mu\nu}$ we recover the usual projective-invariant theory whose connection is the Levi-Civita connection of $h_{\mu\nu}$ up to a projective mode. As a matter of fact, in the generalised RBG case where the projective symmetry is explicitly broken, this projective mode is the only dynamical component of the connection and the remaining components of $\Upsilon$ can be expressed in terms of $B_{\mu\nu}$ by solving (\ref{Eq:Omegaeq}), as we will do later perturbatively up to lowest order in $B^{\mu\nu}$.\\

Since the equations are linear in $\Upsilon^\al{}_{\mu\nu}$, the projective mode can be regarded as a homogeneous solution for $\Upsilon^\alpha{}_{\beta\gamma}$ in the general case, i.e., it belongs to the kernel of \eqref{Eq:Omegaeqdivfree} . In order to isolate this projective mode (homogeneous solution) from the remaining non-dynamical part of the connection (non-homogeneous solution), it is common to introduce the shifted connection
\beq\label{Eq:Omegah}
\Upsilonh^\alpha{}_{\mu\nu}=\Upsilon^\alpha{}_{\mu\nu}+\frac{1}{D-1}\Upsilon_\mu\delta^\alpha{}_\nu
\eeq
with $\Upsilon_\mu=2\Upsilon^\alpha{}_{[\alpha\mu]}$. This shifted connection satisfies $\Upsilonh^\alpha{}_{[\alpha\mu]}=0$ and it is invariant under a projective transformation of $\Upsilon^\al{}_{\mu\nu}$. In terms of these variables the action can be written as
\beq
\begin{split}
\cS=\frac12\int\dd^Dx\sqrt{-h}\Big[R(h)&-\frac{2}{D-1}B^{\mu\nu}\partial_{[\mu}\Upsilon_{\nu]}+\Upsilonh^\alpha{}_{\alpha\lambda}\Upsilonh^\lambda{}_{\kappa}{}^\kappa-\Upsilonh^{\alpha\mu\lambda}\Upsilonh_{\lambda\alpha\mu}\\
&-\Upsilonh^\alpha{}_{\alpha\lambda}\Upsilonh^\lambda{}_{\mu\nu}B^{\mu\nu}-\Upsilonh^\alpha{}_{\nu\lambda}\Upsilonh^\lambda{}_{\alpha\mu}B^{\mu\nu}-B^{\mu\nu}\nabla^h_\alpha\Upsilonh^\alpha{}_{\mu\nu}
-B^{\mu\nu}\nabla^h_\nu\Upsilonh^\alpha{}_{\alpha\mu}+\cU(B)\Big].\label{RBGactionExpanded}
\end{split}
\eeq
We then see that the projective mode $\Upsilon_\mu$ is in fact the responsible for the divergence-free constraint on the 2-form field. From this form of the action we can already understand the root of the pathologies. Firstly, the absence of a pure kinetic term for the projective mode will render this sector unstable on arbitrary $B_{\mu\nu}$ backgrounds. To show this, let us consider a background where the 2-form develops a non-trivial profile. On such a background, and leaving out kinetic terms and/or non-minimal couplings that will not affect our argument here, the relevant sector is described by
\beq
\cS\supset\int\dd^Dx\sqrt{-h}\Big(B^{\mu\nu}\partial_{[\mu}\Upsilon_{\nu]}-m^2M^{\alpha\beta\mu\nu}B_{\alpha\beta}B_{\mu\nu}\Big),
\eeq
where $m^2$ is some mass parameter and $M^{\alpha\beta\mu\nu}$ the mass tensor that depends on the background configuration, with the obvious symmetries of being antisymmetric in the first and second pair of indices and symmetric under the exchange $(\alpha\beta)\leftrightarrow (\mu\nu)$. If the background 2-form field is trivial, the mass tensor reduces to $M^{\alpha\beta\mu\nu}=h^{\alpha[\mu}h^{\nu]\beta}$ so we have
\beq
\cS\supset\int\dd^Dx\sqrt{-h}\Big(B^{\mu\nu}\partial_{[\mu}\Upsilon_{\nu]}-m^2B_{\mu\nu}B^{\mu\nu}\Big).
\eeq
We can diagonalise this sector by performing the field re-definition $B^{\mu\nu}=\hat{B}^{\mu\nu}+\frac{1}{2m^2}\partial^{[\mu}\Upsilon^{\nu]}$, an the above action now reads
\beq
\cS\supset\int\dd^Dx\sqrt{-h}\left(\frac{1}{4m^2}\partial_{[\mu}\Upsilon_{\nu]}\partial^{[\mu}\Upsilon^{\nu]}-m^2\hat{B}_{\mu\nu}\hat{B}^{\mu\nu}\right).
\eeq
Once this sector of the gravitational action has been diagonalised, it becomes apparent that the projective mode acquires the usual gauge-invariant Maxwellian kinetic term for a vector field, but with the wrong sign. One could obtain the correct sign by assuming $m^2<0$, but then the 2-form sector would have the wrong sign for the mass term and, consequently, the ghost would appear there. Either case, we clearly see that the presence of a ghost around a trivial $B^{\mu\nu}$ background is unavoidable. However, there is the possibility that within a non-trivial $B^{\mu\nu}$ background the 2-form field behaves as a ghost condensate. In order to see if this is the case, notice that in a general $B^{\mu\nu}$ background, the diagonalisation requires a field re-definition of the form 
\beq
B^{\mu\nu}=\hat{B}^{\mu\nu}+\frac{1}{2m^2}\Lambda^{\mu\nu\alpha\beta}\partial_{[\alpha}\Upsilon_{\beta]}
\eeq
with $\Lambda^{\mu\nu\alpha\beta}$ satisfying generally
\beq
M^{\alpha\beta\lambda\kappa}\Lambda_{\lambda\kappa}{}^{\mu\nu}=h^{\alpha[\mu}h^{\nu]\beta}.
\label{Eq:LambdaM}
\eeq
In this case, the relevant sector of the gravitational action can be written as
\beq
\cS\supset\int\dd^Dx\sqrt{-h}\left(\frac{1}{4m^2}\Lambda^{\alpha\beta\mu\nu}\partial_{[\alpha}\Upsilon_{\beta]}\partial_{[\mu}\Upsilon_{\nu]}-m^2M^{\alpha\beta\mu\nu}B_{\alpha\beta}B_{\mu\nu}\right).
\eeq
To see whether the ghost persists in general we have to look to the signature character of $\Lambda^{\alpha\beta\mu\nu}$ and $M^{\alpha\beta\mu\nu}$. The ghostly nature of the projective mode is avoided if $\Lambda^{\alpha\beta\mu\nu}$ is a super-metric with the same {\it signature} as $-h^{\alpha[\mu}h^{\nu]\beta}$, being $h^{\mu\nu}$ a Lorentzian metric. On the other hand, stability of the 2-form sector requires a mass tensor with the {\it signature} of $h^{\alpha[\mu}h^{\nu]\beta}$. These two conditions are however inconsistent with each other by virtue of the relation (\ref{Eq:LambdaM}) and therefore no ghost condensation can stabilise the theory. Thus, we find that the presence of a ghost in the projective sector of generalised RBGs is unavoidable and occurs in an arbitrary background. This is the ghost found in \ref{sec:InstDecLim} beyond the decoupling limit. \\

It is interesting to notice that the re-definition of the 2-form field that diagonalises the quadratic action for the trivial background configuration corresponds to a gauge-like transformation for the 2-form, hence, its field strength will be oblivious to such re-definition. In particular, this means that kinetic terms with the correct gauge invariant form $H^2$ will not be affected by the diagonalisation and, therefore, cannot change our conclusion about the presence of a ghost. The same reasoning applies to non-trivial backgrounds that vary weakly as compared to $m^2$. If this is not the case, one might envision that sufficiently strongly varying backgrounds could give rise to a stabilisation \`a la ghost condensate. Even without taking into account couplings to gravity, it should be apparent that there will always be UV modes with a sufficiently high frequency for which the background is effectively constant and, therefore, our discussion above will also apply, thus showing the pathological character of these modes. A natural way around this problem is to assume that those modes are beyond the regime of validity of the theory and, consequently, it does not pose an actual problem. In that case however, the full EFT approach should be taken from the very beginning. Moreover, there will also be non-minimal couplings to the curvature, which after diagonalisation will introduce yet additional pathologies arising from that sector so our hopes stand on shaky grounds anyways. To understand this, we must look at the connection equations \eqref{Eq:Omegaeq}, from where it is apparent that the solution for $\Upsilonh$ will have the schematic form
\beq
\Upsilonh\sim \frac{\nabla^h B}{1+B}.
\eeq
Plugging this solution back into the RBG action written as \eqref{RBGactionExpanded} and integrating out the non-dynamical piece of the connection $\Upsilonh$, additional terms like $(\nabla^h B)^2$ and $B(\nabla^h)^2B$ will arise. The latter can be integrated by parts to be put in the form of the former. Doing this however can result in non-gauge invariant derivative terms and/or non-minimal couplings arising from commuting covariant derivatives. Both of such terms are potentially dangerous and the source of ghost-like instabilities. It is remarkable that the quadratic derivative terms generated in the action can be brought into the standard gauge-invariant kinetic term of a two form. However, this is an accident of the leading order solution and it is broken at higher orders. Let us see this explicitly.

%%%%%%%%%%%%%%%%%%%%%%%%%%%%%%%%%%%%%%%%%%%%%%%%%%%%%%%%%%%%%%%%%%%%%%%%%%%%%%%%%%%%%%%%%%%%%%%%%%%%%%%%%%%%%%%%%%%%%%%%%%%%%%%%%%%%%%%%%%%%%%%%%%%%%%%%%%%%%%%%%%%%%%%%%%%%%%%%%%%%

\subsection{Solving for the connection}\label{sec:SolConn}
We will illustrate the form of the solutions for the connection by considering vacuum configurations, so that the action is given by
\beq
\cS=\frac12\int \dd^Dx \Big[\sqrt{-q}q^{\mu\nu}\cR_{\mu\nu}(\Ga)+\cU(q)\Big].
\eeq
The connection equations for this action are the same as we obtained in \eqref{conneq} or (\ref{Eq:Gammah}), i.e., the connection deprived of its projective mode satisfies
\beq
\partial_\la(\sqrt{-q}q^{\mu\nu})+\hat{\Gamma}^{\mu}{}_{\la\al}\sqrt{-q}q^{\al\nu}+\hat{\Gamma}^{\nu}{}_{\al\la}\sqrt{-q}q^{\mu\al}-\hat{\Gamma}^{\al}{}_{\la\al}\sqrt{-q}q^{\mu\nu}=0.\label{conneq2}
\eeq
this equation does allow, at least formally, to algebraically solve for the connection in terms of $q^{\mu\nu}$. With this aim let us again decompose the connection as in \eqref{Eq:GammasplittingChO}, so that we extract the Levi-Civita connection of the symmetric component of $h^{\mu\nu}$. The projectively transformed connection is therefore given by 
\beq
\hat\Gamma^\alpha{}_{\mu\beta}=\Ch{\alpha}{\mu}{\beta}(h)+\Upsilonh^\alpha{}_{\mu\beta}
\label{Eq:GammasplittingChO1}
\eeq
where $\Upsilonh$ is defined as in 
\eqref{Eq:Omegah}. We can now introduce the above splitting \eqref{Eq:GammasplittingChO1} into the connection equations equations \eqref{conneq2}. By performing the usual trick of adding and subtracting the resulting equation with suitably permuted indices, we can write a formal solution for the connection as
\begin{equation}\label{formalsolconn}
    \Upsilonh^{\alpha}{}_{\mu\nu}=\left[\frac{1}{2} h^{\kappa \lambda}\left(\nabla_{\beta}^{h} B_{\ga \lambda}+\nabla_{\ga}^{h} B_{\lambda \beta}-\nabla_{\lambda}^{h} B_{\beta \ga}\right)\right]\left(A^{-1}\right)_{\kappa}{}^{\alpha}{}_{\mu\nu}{}^{\beta \gamma},
\end{equation}
where by definition ${A}^\kappa{}_{\al'}{}^{\mu'\nu'}{}_{\be\ga} ({A}^{-1})_\kappa{}^\al{}_{\mu\nu}{}^{\be\ga}\equiv \delta^{\al'}{}_{\al} \delta^{\mu'_{\mu}} \delta^{\nu'_{\nu}}$. Here ${A}^\kappa{}_{\al'}{}^{\mu'\nu'}{}_{\be\ga}$ is linear in $B_{\mu\nu}$ and is given by
\begin{align}\label{Operatorsolcon}
\begin{split}
 &{A}^\kappa{}_{\al}{}^{\mu\nu}{}_{\be\ga} \equiv a^\kappa{}_{\al}{}^{\mu\nu}{}_{\be\ga}+b^\kappa{}_{\al}{}^{\mu\nu}{}_{\be\ga}{}^{\rho\sigma}B_{\rho\sigma}\\
 &a^\kappa{}_{\al}{}^{\mu\nu}{}_{\be\ga} \equiv
 \delta^\kappa{}_\al \delta^\mu{}_\be \delta^\nu{}_\ga +\frac{1}{2} \delta^\mu{}_{\alpha}\left(h^{\nu\kappa} h_{\beta \gamma}-2 \delta^{\nu}{}_{(\beta} \delta^\kappa{}_{\gamma)}\right)\\
 &b^\kappa{}_{\al}{}^{\mu\nu}{}_{\be\ga}{}^{\rho\sigma}=\frac{1}{2}\left[h_{\alpha\gamma} h^{\mu\sigma} \delta^\nu{}_{\beta} h^{\rho\kappa}+\delta^{\beta}{}_\rho h_{\alpha\gamma} h^{\mu\kappa} h^{\nu \sigma}+\delta^{\rho}{}_\gamma \delta^\mu{}_{\alpha} \delta^\nu{}_{\beta} h_{\kappa \sigma}-h^{\rho\kappa} \delta^{\mu}{}_\gamma h^{\nu\sigma} h_{\alpha \beta}-\delta^{\rho}{}_ \beta h^{\sigma\kappa} \delta^\mu{}_{\alpha}\delta^{\nu}{}_\gamma\right.\\
&\hspace{2.4cm}\left.-\delta^{\rho}{}_ \beta \delta^{\sigma}{}_\gamma\delta^{\mu}{}_\alpha h^{\nu\kappa}-\delta^{\rho}{}_\gamma h_{\alpha \beta} h^{\mu{\sigma}} h^{\nu{\alpha}}-\delta^{\rho}_{\gamma} \delta^{\kappa}{}_{\alpha} \delta^{\mu}{}_\beta h^{\nu\sigma}+\delta^{\rho}{}_\beta \delta^{\kappa}{}_ \alpha h^{\mu \sigma} \delta^{\nu}{}_\gamma\right].
\end{split}
\end{align}
In order to explicitly show the appearance of problematic couplings, it will suffice to give a perturbative solution to lowest-order in $B$. To that end, let us consider a trivial 2-form background and expand around it, leaving the symmetric sector $h^{\mu\nu}$ completely general. The only task then is either to compute the $\mathcal{O}(B^0)$ term of $({A}^{-1})_\kappa{}^\al{}_{\mu\nu}{}^{\be\ga}$ or to directly solve the equations \eqref{Eq:Omegaeq} for $\Upsilonh$ expanded as a power series. Let us proceed with the second method by expanding $\Upsilonh$ as a power series of the 2-form $B$ in the form
\beq\label{eq:seriesom}
\Upsilonh^{\al}{}_{\mu\nu}=\sum_{n=0}^{\infty}\Upsilonh_{(n)}{}^\al{}_{\mu\nu},
\eeq
where the sub-index $n$ implies that the quantity is of order $\cO(B^n)$. We can now use \eqref{Eq:GammasplittingChO1} to split the connection symbols that appear in \eqref{conneq2}, and plugging the expansion of $\Upsilonh^{\al}{}_{\mu\nu}$ into the resulting equation, we obtain
\beq
\begin{split}
\na^h_\la B^{\mu\nu}-B^{\mu\nu}\Upsilonh_{(0)}{}^{\al}{}_{\la\al}-B^\nu{}_\al\Upsilonh_{(0)}{}^{\mu\al}{}_{\la}+B^\mu{}_\al\Upsilonh_{(0)}{}^{\nu}{}_\la{}^\al&+h^{\mu\nu}\Upsilonh_{(1)}{}^\al{}_{\la\al}-\Upsilonh_{(1)}{}^{\mu\nu}{}_\lambda-\Upsilonh_{(1)}{}^{\nu}{}_\la{}^\mu-\\&-\Upsilonh_{(0)}{}^{\mu\nu}{}_\la-\Upsilonh_{(0)}{}^{\nu}{}_\lambda{}^\mu+h^{\mu\nu}\Upsilonh_{(0)}{}^{\al}{}_{\la\al}=\cO(B^2).\label{conneqpert}
\end{split}
\eeq
Notice that this equation is consistent with substituting the perturbative series \eqref{eq:seriesom} in \eqref{Eq:Omegaeqdivfree}, as it should be.\footnote{To see this explicitly, one should take into account the equation resulting from contracting $\alpha$ and $\mu$ in  \eqref{Eq:Omegaeqdivfree} together with the identity $\Upsilonh^\al{}_{[\al\be]}=0$, wich leads to $\Upsilonh^{\nu\beta}{}_{\beta}- B_{\beta\lambda}\Upsilonh^{\nu\beta\lambda}=0$.} The zeroth order term gives the equation
\beq
\Upsilonh_{(0)}{}^{\mu\nu}{}_\la+\Upsilonh_{(0)}{}^{\nu}{}_\lambda{}^\mu-h^{\mu\nu}\Upsilonh_{(0)}{}^{\al}{}_{\la\al}=0,
\eeq
which after contracting with $h_{\mu\nu}$ gives $\Upsilonh_{(0)}{}^\al{}_{\la\al}=0$ for $D\neq2$. This leaves us with the equation $\Upsilonh_{(0)}{}^{\mu\nu}{}_\la+\Upsilonh_{(0)}{}^{\nu}{}_\lambda{}^\mu=0$. Doing the permutating trick we arrive to 
\beq
\Upsilonh_{(0)}{}^\al{}_{\mu\nu}=0,
\eeq
which ensures that the Levi-Civita connection of $h^{\mu\nu}$ is the solution (up to a projective mode) for the affine connection for a symmetric metric. Note that this was expected since for vanishing $B^{\mu\nu}$ we are describing GR, which has the Levi-Civita connection of the metric as the only solution (up to a projective mode) \cite{Bernal:2016lhq,Janssen:2019htx}. Plugging the 0th order result into \eqref{conneqpert}, we arrive to the equation for the $\cO(B)$ piece of $\Upsilonh^\al{}_{\mu\nu}$, which reads
\beq
\na^h_\la B^{\mu\nu}-\Upsilonh_{(1)}{}^{\mu\nu}{}_\la-\Upsilonh_{(1)}{}^\nu{}_\la{}^\mu-h^{\mu\nu}\Upsilonh_{(1)}{}^\al{}_{\la\al}=0.
\eeq
Contracting with $h_{\mu\nu}$ we the condition  $\Upsilonh_{(1)}{}^\al{}_{\la\al}=0$ for $D\neq2$, which leads to the equation $\na^h_\la B^{\mu\nu}-\Upsilonh_{(1)}{}^{\mu\nu}{}_\la-\Upsilonh_{(1)}{}^{\nu}{}_\lambda{}^\mu=0$. This can be solved again by performing the permutation trick, thus obtaining
\beq\label{pertsolcon}
\Upsilonh_{(1)}{}^\al{}_{\mu\nu}=\frac{1}{2}h^{\al\la}\left(\na^h_\mu B_{\nu\la}+\na^h_\nu B_{\la\mu}-\na^h_\la B_{\mu\nu}\right),
\eeq
in agreement with previous results in NGT obtained in \cite{Damour:1991ru} and the formal solution \eqref{formalsolconn} given above\footnote{Although the form of $a^\kappa{}_{\al}{}^{\mu\nu}{}_{\be\ga}$ in \eqref{Operatorsolcon} suggests than the formal solution has more contributions to first order in $B$ than \eqref{pertsolcon}, it can be seen that ${a}^\kappa{}_{\al}{}^{\mu\nu}{}_{\be\ga}\hat\Upsilon_{(1)}{}^{\al}{}_{\mu\nu}=\Upsilon_{(1)}{}^{\kappa}{}_{\be\ga}+\mathcal{O}(B^2)$, which implies that the formal solution \eqref{formalsolconn} and the first-order perturbative one \eqref{pertsolcon} are consistent.}. As stated in the end of the previous section, and analogously to the results on NGT in \cite{Damour:1991ru}, the dependence of $\Upsilonh$ on the derivatives of $B^{\mu\nu}$ will introduce additional pathologies in the 2-form field. As a matter of fact, upon substitution of this solution into \eqref{RBGactionExpanded} and integration by parts, we arrive at the desired action similar to \eqref{eq:Actquad} featuring a gauge-invariant kinetic term for the 2-form together with the non-minimal couplings advertised above. Again, the gauge invariance of the derivative operators for the 2-form is accidental of this order, but it is broken at cubic and higher orders. It is possible, although tedious, to obtain the solution for $\Upsilon$ at arbitrary order by following this perturbative scheme. Obtaining a full solution in closed form appears to be a more challenging task.

\subsection{On more general actions}\label{sec:OnMoreGeneral}

So far we have focused on theories constructed in terms of the Ricci tensor alone as a simplified proxy to prove the pathological character of general metric-affine theories described by higher order curvature actions. Our results should suffice to clearly identify the origin for the potential pathologies in more general metric-affine theories where not only the Ricci tensor appears in the action, but arbitrary non-linear terms constructed with the Riemann curvature tensor. In general, if we have an action with an arbitrary dependence on the Riemann tensor formulated in an affine geometry, we can always introduce the splitting of the connection into its Levi-Civita part, the torsion and the non-metricity. That way, it is possible to re-formulate the theory in a pure Riemannian geometry with additional fields. These fields, i.e. the torsion and the non-metricity, can be decomposed into their irreducible representations under some appropriate group, GL(4,$\mathbbm{R}$) or ISO(1,3) being natural choices (see e.g. \cite{Hehl:1994ue}), and they will feature non-minimal couplings to the curvature and, quite generically, these will involve either derivatives of the fields or couplings of spin higher than zero. In both cases, as it is well-known, such interactions are prone to pathologies, specially to Ostrogradski instabilities. In the precedent sections we have explicitly shown how these expected pathologies come about for a particular class of theories and only when the extra fields drop from the spectrum can we have stable theories, in which case they simply reduce to GR, but it is clear that the same problems will persist for more general actions. 

It is important to emphasise that we are providing a general argument against some commonly quoted statements\footnote{From a field theory perspective it is evident that having second order field equations is not a sufficient reason to guarantee the absence of Ostrogradski instabilities, a straightforward argument being that it is always possible to reduce the order of the equations by introducing auxiliary fields. However, in the community with a stronger geometrical approach to gravity this seems to be less clear.} that the metric affine theories avoid instabilities because the field equations remain of second order. This does not mean however that {\it all} metric-affine theories with higher order curvature terms featuring additional propagating dofs (other than the graviton) will be pathological, but one should be careful on how these theories are constructed and not give for granted that the very fact of using a metric-affine formulation prevents the appearance of ghosts from operators involving arbitrary powers of the Riemann tensor. Of course, non-pathological theories exist and they can be constructed in a variety of manners (some of which we will discuss in Sec. \ref{sec:GeoConst}), usually introducing additional symmetries, constraints or geometrical identities. However, it should be clear from our discussion that one should be careful when constructing theories in a metric-affine framework. The general problematic character of metric-affine gravity theories can be seen from the analysis of the perturbative dof's around Minkowski performed in \cite{Percacci:2019hxn} where it was shown that, already at that level, wise choices of parameters must be taken to avoid instabilities. It is important however to stress that our analysis above goes beyond the linear regime around Minkowski and, in fact, some of the diagnosed instabilities cannot be seen from such a perturbative analysis. Thus, though the perturbative analysis gives necessary conditions for stability, it is not sufficient to ensure the non-linear stability of the theories. For an example of how the perturbative analysis is not sufficient see e.g. \cite{Yo:1999ex,Jimenez:2019qjc,Jimenez:2019tkx} within the context Poincar\'e gauge theories.

Let us finally briefly comment on how our results can be relevant for a pure effective field theory (EFT) approach to the metric-affine theories theories. This approach has been thoroughly pursued in \cite{Aoki:2019snr} within the class of Riemann-Cartan geometries including up to dimension 4 operators. We have seen that higher order powers of the Riemann tensor generically introduce ghosts-like instabilities in the metric-affine formalism very much like in the metric approach (essentially for the same reasons). It is possible however to adopt an EFT approach where these will just be irrelevant operators with perturbative effects below the cut-off of the theory. In this view, the ghosts are not really part of the perturbative spectrum of the theory because their masses are beyond the cut-off scale so they are harmless. If the gravitational cut-off  is assumed to be the Planck mass and the Wilsonian coefficients are $\mathcal{O}(1)$ according to naturalness arguments, then the EFT is will be similar (with additional fields) to the usual EFT approach to GR. On the other hand, if we assume that the Planck scale only represents the cut-off for the purely metric sector and the metric-affine sector comes in with another scale $M<\mpl$, then one would expect the EFT theory to breakdown at that scale. This implies for instance that classical solutions where the curvature becomes larger than $M$ cannot be generically trusted.

%%%%%%%%%%%%%%%%%%%%%%%%%%%%%%%%%%%%%%%%%%%%%%%%%%%%%%%%%%%%%%%%%%%%%%%%%%%%%%%%%%%%%%%%%%%%%%%%%%%%%%%%%%%%%%%%%%%%%%%%%%%%%%%%%%%%%%%%

\section{Matter couplings}\label{sec:Mattercoupling}

In the precedent sections we have only considered matter fields which do not couple to the connection. However, our conclusions on the presence of pathological dof's  do not change substantially by coupling the connection to the matter sector. Couplings to matter fields in a metric-affine framework is an interesting issue by itself, specially when it involves spinor fields (see e.g. \cite{Hehl:1994ue,Delhom:2020hkb,BeltranJimenez:2020sih}). It is not the scope of this section to carefully go through the different coupling prescriptions to matter nor their consistency. Our aim is to show how our results above are not substantially affected in the presence of matter fields with minimal couplings as well as discussing some non-minimal couplings that can be safely included. 

\subsection{The Non-symmetric gravity frame for non-minimally coupled fields}\label{sec:NGTframemat}

Curvature couplings to the matter sector include derivatives of the affine connection in the matter Lagrangian. This further complicates the connection equation \eqref{GeneralisedConneq} by adding extra terms on the right hand side. However, there is a class of couplings for which, while adding technical complications, the qualitative results remain the same with just some minor adjustments with respect to the minimally coupled fields. We will start by considering bosonic fields whose non-minimal couplings are through the Ricci tensor. To illustrate this point, we can consider a scalar field $\varphi$ as a proxy for the matter sector. If we restrict to only first derivatives of the scalar, we can use for instance $\cR^{\mu\nu}\partial_\mu\varphi\partial_\nu\varphi$ or $\cR(\partial\varphi)^2$ in our action. In the usual metric formalism, these two terms are only allowed if they enter through the specific combination $(R^{\mu\nu}-\frac12 Rg^{\mu\nu})\partial_\mu\varphi\partial_\nu\varphi$ and accompanied by the appropriate second derivative interactions of the scalar field in order to avoid Ostrogradski instabilities. In the metric-affine formalism however, this is not necessary and the dependence on said terms is completely arbitrary. Let us note that these interactions will not break the projective symmetry since they only depend on the symmetric part of the Ricci tensor. Interestingly, it has been suggested in \cite{Aoki:2019rvi} that the projective symmetry could also play a crucial role to guarantee the absence of ghosts for theories containing up to second order covariant derivatives of a scalar field. They also find that in the stable theories the connection is devoid of any propagating mode as a consistency condition as we argued above.
\\

Our reasoning can be straightforwardly extended to other fields such as vector fields $A_\mu$ where interactions like $\cR_{\mu\nu}A^\mu A^\nu$ or $\cR_{\mu\nu} F^{\mu\alpha}F^\nu{}_\alpha$ also respect the projective symmetry and are permitted. The crucial point of all these interactions is that an Einstein frame still exists where it is apparent that the connection remains an auxiliary field \cite{Afonso:2017bxr}. In the absence of the projective symmetry, we will encounter the same pathologies as exposed for the pure gravitational sector and the inclusion of a contrived matter sector cannot remedy it. In Sec. \ref{sec:NSGFrame} we showed how to go to the Einstein frame of RBG theories for minimally coupled matter fields. Let us see here how to proceed in the presence of non-minimally coupled matter fields. In this case the action reads
\beq\label{GeneralActionHyp}
\cS[g_{\mu\nu},\Ga,\Psi]=\frac12\int \dd^D x \sqrt{-g}\,F\big(g^{\mu\nu},\cR_{\mu\nu}\big)+\cS_m[g,\Psi,\Gamma].
\eeq
Parallel to \ref{sec:ProblemAdditionalDOFs}, we now go to the Einstein frame of the above theory, and after splitting the corresponding auxiliary metric as in \eqref{metricsplitting} and the connection as in \eqref{Eq:GammasplittingChO}, and also isolating the projective mode from $\Upsilon^\al{}_{\mu\nu}$ as in \eqref{Eq:Omegah}, we get 

\beq
\begin{split}
\cS=\frac12\int\dd^Dx\sqrt{-h}\Big[R(h)&-\frac{2}{D-1}B^{\mu\nu}\partial_{[\mu}\Upsilon_{\nu]}+\Upsilonh^\alpha{}_{\alpha\lambda}\Upsilonh^\lambda{}_{\kappa}{}^\kappa-\Upsilonh^{\alpha\mu\lambda}\Upsilonh_{\lambda\alpha\mu}-\Upsilonh^\alpha{}_{\alpha\lambda}\Upsilonh^\lambda{}_{\mu\nu}B^{\mu\nu}\\
&-\Upsilonh^\alpha{}_{\nu\lambda}\Upsilonh^\lambda{}_{\alpha\mu}B^{\mu\nu}-B^{\mu\nu}\nabla^h_\alpha\Upsilonh^\alpha{}_{\mu\nu}
-B^{\mu\nu}\nabla^h_\nu\Upsilonh^\alpha{}_{\alpha\mu}+\cU(B)\Big]+\tilde\cS_m[h,B,\Psi,\hat\Upsilon,\Upsilon].\label{RBGactionExpandedHyp}
\end{split}
\eeq
where now $\tilde\cS_m$ is the matter action in the Einstein frame, and the variables inside square brackets means that the matter action can depend on those fields and their derivatives in general. Concretely $\Upsilon$ stands for the dependence of the matter action on the projective mode, so it will be absent for projectively invariant matter. It is then apparent that the gravitational sector features the same pathological terms. Obviously, a trivial matter sector background will not modify those terms. A non-trivial matter background contributing to the background symmetric part of the metric could help by providing a kinetic term for the projective mode. However, the non-minimal couplings to the curvature for the 2-form that are generated after integrating $\hat{\Upsilon}$ out are hardly cured. In any case, this would require very specific choices of the matter sector. To make this statement more explicit, let us consider a particular class of matter sector coupled to the connection.

\subsection{Ultra-local matter couplings}\label{sec:SolConmat}

For mater actions which do not include curvature couplings (i.e. no derivatives of the connection), we already know that the projective mode will be problematic due to the absence of a proper kinetic term for it. In order to understand if the inclusion of a general coupling between matter and connection can solve the instability problems we can now compare the above action \eqref{RBGactionExpandedHyp} to \eqref{RBGactionExpanded}. First notice that the divergence-free constraint of the 2-form \eqref{DivConstB} that came from the field equations of the projective mode gets modified if non-projectively invariant matter actions are taken into account, and the trace of the hypermomentum acts now as a source for $B$
\beq
\na^h_\mu B^{\mu\nu}=\frac{D-1}{4}\Delta_\al{}^{[\mu\al]}
\eeq

where $\Delta_{\lambda}^{\mu \nu}$ is the hypermomentum defined as

\begin{equation}\label{DefHypermomentum}
\left.\Delta_{\lambda}{}^{\mu \nu} \equiv 2 \frac{\delta \mathcal{S}_{m}}{\delta \Gamma^{\lambda}{}_{\mu \nu}}\right|_{g_{\mu \nu}}=2 \left.\frac{\delta \mathcal{S}_{m}}{\delta \Upsilon^{\lambda}{}_{\mu \nu}}\right|_{g_{\mu \nu}}
\end{equation}
which vanishes for matter fields that do not couple to the connection. Looking at the form of this action, we can see that the projective mode will in general feature the same problems as in the previous case when the matter and connection did not couple. The Ostrogradski instabilities that arise from the couplings between the 2-form $B^{\mu\nu}$ and the curvature of $h_{\mu\nu}$ will still be there no matter what matter action we choose. Therefore, we see that allowing for an arbitrary coupling between matter and connection is not helpful in solving any of the instabilities listed above. To explicitly see what kind of couplings arise, we have to solve the connection equation now with a hypermomentum. Since generally an analytic solution is not possible, and even if it is, it is not very illuminating, we will attempt to find a perturbative solution which will already give us a clear picture of the issue. Let us then write down the connection equations when a coupling between matter and connection is present:
\begin{equation}\label{GeneralisedConneq}
\begin{split}
{\nabla_{\lambda}\left[\sqrt{-q} q ^{\nu\mu}\right]-\del^\mu{}_\la \nabla_{\rho}\left[\sqrt{-q} q ^{ \nu\rho}\right]}{=\Delta_{\lambda}{}^{\mu \nu}+\sqrt{-q}\left[\mathcal{T}^{\mu}{}_{\lambda \alpha} q ^{\nu\al}+\mathcal{T}^{\alpha}{}_{\alpha \lambda} q ^{\nu\mu}-\delta_{\lambda}^{\mu} \mathcal{T}^{\alpha}{}_{\alpha \beta} q ^{\nu\be}\right]}.
\end{split}
\end{equation}
  In order to remain as close as possible to the previous analysis in sec.\ref{sec:SolConn}, it is necessary to use the shifted connection \eqref{Eq:Gammah} and find the relation between the hypermomentum of the original connection $\Delta_\al{}^{\mu\nu}$ and the shifted hypermomentum  $\Deltah_\al{}^{\mu\nu}$, which reads
\beq
\Delta_\al{}^{\mu\nu}=\Deltah_\al{}^{\mu\nu}+\frac{2}{D-1}\delta_\al{}^{[\mu}\Deltah_\be{}^{\nu]\be},
\eeq
where the shifted hypermomentum is defined in an analogous manner as \eqref{DefHypermomentum}. This implies that the hyermomentum of projectively invariant matter fields satisfies $\Deltah_\be{}^{\mu\be}$. We can now recast \eqref{GeneralisedConneq} in the form of \eqref{conneq} by doing the same manipulations, thus finding
\beq
\partial_\la(\sqrt{-q}q^{\mu\nu})+\hat{\Gamma}^{\mu}{}_{\la\al}\sqrt{-q}q^{\al\nu}+\hat{\Gamma}^{\nu}{}_{\al\la}\sqrt{-q}q^{\mu\al}-\hat{\Gamma}^{\al}{}_{\la\al}\sqrt{-q}q^{\mu\nu}=\Deltah_\al{}^{\mu\nu}+\frac{2}{D-1}\delta_\al{}^{[\mu}\Deltah_\be{}^{\nu]\be}.\label{conneqhyperm}
\eeq
 As in the vanishing hypermomentum case, we can obtain a formal solution for the full connection in the case of arbitrary hypermomentum as
\begin{equation}
    \Upsilonh^{\alpha}{}_{\mu\nu}=\frac{1}{2} h^{\kappa \lambda}\left[\left(\nabla_{\beta}^{h} B_{\ga \lambda}+\nabla_{\ga}^{h} B_{\lambda \beta}-\nabla_{\lambda}^{h} B_{\beta \ga}\right)+\frac{1}{\sqrt{-h}} h^{\kappa \lambda}\left(\Deltah_{\beta\ga \lambda}+\Deltah_{\ga\lambda \beta}+\Deltah_{\lambda\beta \ga}+\frac{2}{D-1}h_{\lambda[\gamma}\Deltah^\al{}_{\be]\al}\right)\right]\left(A^{-1}\right)_{\kappa}{}^{\alpha}{}_{\mu\nu}{}^{\beta \gamma},
\end{equation}
where $\left(A^{-1}\right)_{\kappa}{}^{\alpha}{}_{\mu\nu}{}^{\beta \gamma}$ is the same operator as in the vanishing hypermomentum case, which is specified in \eqref{Operatorsolcon}. Notice that the above formula points to the fact that the addition of hypermomentum does not solve any of the instabilities due to the dependence of $\Upsilonh$ on the derivatives of $B_{\mu\nu}$. To see that this is the case, let us find a perturbative solution to the connection in an analogous way to that of \ref{sec:SolConn}. First we need to write $\Deltah_ \al{}^{\mu\nu}=\Deltah^{(0)}_ \al{}^{\mu\nu}+\Deltah^{(1)}_ \al{}^{\mu\nu}+...$ as a power series in $B^{\mu\nu}$, where the superscript $(n)$ indicates that such term is of order $\mathcal{O}(B^n)$. Then, after splitting the shifted connection as in \eqref{Eq:GammasplittingChO1} and then writing $\Upsilonh^\al{}_{\mu\nu}$ as a power series in $B^{\mu\nu}$ as in \eqref{eq:seriesom}, we can write \eqref{GeneralisedConneq} in an analogous fashion to \eqref{conneqpert} as
\beq
\begin{split}
\na^h_\la B^{\mu\nu}-B^{\mu\nu}&\Upsilonh_{(0)}{}^{\al}{}_{\la\al}-B^\nu{}_\al\Upsilonh_{(0)}{}^{\mu\al}{}_{\la}+B^\mu{}_\al\Upsilonh_{(0)}{}^{\nu}{}_\la{}^\al+h^{\mu\nu}\Upsilonh_{(1)}{}^\al{}_{\la\al}-\Upsilonh_{(1)}{}^{\mu\nu}{}_\lambda-\Upsilonh_{(1)}{}^{\nu}{}_\la{}^\mu-\Upsilonh_{(0)}{}^{\mu\nu}{}_\la\\
-&\Upsilonh_{(0)}{}^{\nu}{}_\lambda{}^\mu+h^{\mu\nu}\Upsilonh_{(0)}{}^{\al}{}_{\la\al}-\Deltah^{(0)}_\al{}^{\mu\nu}+\frac{2}{D-1}\delta_\al{}^{[\mu}\Deltah^{(0)}_\be{}^{\nu]\be}-\Deltah^{(1)}_\al{}^{\mu\nu}+\frac{2}{D-1}\delta_\al{}^{[\mu}\Deltah^{(1)}_\be{}^{\nu]\be}=\cO(B^2).
\end{split}
\eeq
 Notice that in general, $\Deltah^{(n)}_\al{}^{\mu\nu}$ might have a complicated dependence on the affine connection, and thus on $\Upsilonh^\al{}_{\mu\nu}$, which may complicate further the solution of the above equation for $\Upsilonh^\al{}_{\mu\nu}$ order by order in $B^{\mu\nu}$. Thus, in general, one could make a further expansion of each $\Deltah^{(n)}_\al{}^{\mu\nu}=\Deltah^{(0,n)}_\al{}^{\mu\nu}+\Deltah^{(1,n)}_\al{}^{\mu\nu}+...$ where the superscript $(m,n)$ denotes a term of order $\mathcal{O}(\Upsilonh^m)$ and $\mathcal{O}(B^n)$. Since the completely general case is rather cumbersome, and is not particularly illuminating, let us focus on the case where the hypermomentum does not depend on the affine connection, where we can expand only in terms of $B^{\mu\nu}$. Let us mention that this would be the case, for instance, of minimally coupled spin 1/2 fields, which have a well known hypermomentum of the form $\Delta^{(\Psi)}_\al{}^{\mu\nu}=-i\sqrt{-q}g_{\al\rho}\epsilon^{\rho\sigma\mu\nu}\lrsq{\bar{\Psi}\gamma_\sigma\gamma_5\Psi}.$ Assuming thus no dependence of the hypermomentum on the connection\footnote{The equation of zeroth order would still be formally valid for $\hat\Delta_{\al}{}^{\mu\nu}$ that depends on the connection, altough in that case it will be harder to issolate $\Upsilon^\al{}_{\mu\nu}$.} (i.e. the matter action is linear in the connection), we can proceed exactly as in sec.\ref{sec:SolConn} to obtain the following zeroth and first order solutions:
\beq
\begin{split}
\Upsilonh_{(0)}{}^\al{}_{\mu\nu}=&\frac{1}{\sqrt{-h}}\lrsq{\Deltah^{(0)}{}_{\mu\nu}{}^\al+\Deltah^{(0)}_\nu{}^\al{}_\mu-\Deltah^{(0)}{}^\al{}_{\mu\nu}+\frac{2}{D-1}\delta^\la{}_{[\nu}\Deltah^{(0)}{}^\al{}_{\mu]\la}+\frac{1}{2(D-2)}\lr{h_{\mu\nu}\Deltah^{(0)}{}^{\al\la}{}_\la-2\delta^\al{}_{(\mu}\Deltah^{(0)}{}_{\nu)\la}{}^\la}}\\
\Upsilonh_{(1)}{}^\al{}_{\mu\nu}=&\frac{1}{2}h^{\al\la}\Big(\na^h_\la B_{\nu\mu}+\na^h_\mu B_{\nu\la}+\na^h_\nu B_{\la\mu}\Big)-\frac{1}{\sqrt{-h}}\Bigg[\frac12\lr{\Deltah^{(1)}{}^\al{}_{\mu\nu}+\Deltah^{(1)}{}_{\mu\nu}{}^\al+\Deltah^{(1)}{}_\nu{}^\al{}_{\mu}}-\frac{1}{D-2}\delta^\al{}_{(\mu}\Deltah^{(0)}{}_{\nu)}{}^{\ga\sigma}B_{\ga\sigma}\\
&+\frac{2}{(D-1)(D-2)}\delta^\al{}_{(\mu}B_{\nu)\ga}\Deltah^{(0)}{}_\sigma{}^{\sigma\ga}\frac{1}{D-2}\delta^\al{}_{(\mu}\Deltah^{(1)}{}_{\nu)\sigma}{}^{\sigma}+\frac{1}{D-2}\delta^\al{}_{[\mu}B_{\nu]\ga}\Deltah^{(0)}{}^{\ga\sigma}{}_\sigma+\frac{2}{D-1}\delta^\al{}_{[\mu}\Deltah^{(1)}{}^{\sigma}{}_{\nu]\sigma}\\
&+\frac12\lr{ B_{\mu\sigma}\Deltah^{(0)}{}^{\sigma}{}_{\nu}{}^{\al}-B_{\nu\sigma}\Deltah^{(0)}{}^{\sigma\al}{}_{\nu}}+\frac{1}{2(D-2)}h_{\mu\nu}B_{\ga\sigma}\Deltah^{(0)}{}^{\al\ga\sigma}-\frac{1}{D-2}h_{\mu\nu}B^\al{}_\sigma\Deltah^{(0)}{}_{\ga}{}^{\ga\sigma}-\frac{1}{2(D-2)}h_{\mu\nu}\Deltah^{(1)}{}^{\al\sigma}{}_{\sigma}\Bigg],
\end{split}
\eeq
where $\Deltah^{(0)}{}_\al{}^{(\al\be)}=0$ and $\Upsilonh_{(0)}{}^{\al\be}{}_{\be}=0$ must be satisfied as can be shown from the connection field equations and the identity $\Upsilonh_{(n)}{}^\al{}_{[\al\be]}=0$. As we can see, besides obtaining the problematic $\Upsilonh\sim\na^h B+\mathcal{O}(B^2)$ terms that we obtained in the vanishing hypermomentum case, we here obtain also a bunch of terms that couple non-minimally the matter fields with themselves and with the 2-form $B^{\mu\nu}$ through their hypermomentum. It is apparent that the addition of these new terms cannot heal the problematic behaviour of the $\na^h B$ terms by themselves, thus clarifying why the addition of non-minimal couplings to matter fields would not solve the instability problem. Indeed, the extra couplings between the unstable 2-form and the matter fields potentially reduce the time-scale in which the 2-form instability manifests physically through its decay to lighter particles.

\subsection{A digression on metric vs affine geodesic equation}\label{sec:geodesics}

After discussing the consequences of ultra local couplings between the matter fields and the affine connection, let us take on the discussion about the propagation of test particles initiated in Sec. \ref{sec:Generalities}. We indicated there how the projective symmetry was related to the re-parameterisation invariance of the autoparallel or affine geodesic equation \eqref{pregeodesiceq} that for affinely parameterised curves can be written as
 \beq
 \ddot{x}^\alpha+\Gamma^\alpha{}_{\mu\nu}\dot{x}^\mu\dot{x}^\nu=0.
 \label{eq:autoparallel1}
 \eeq
This equation describes the {\it straightest} paths defined as those whose acceleration along the tangent direction vanishes, while the shortest paths are described by the metric geodesic equation
\beq
\ddot{x}^\alpha+\bar{\Gamma}^\alpha{}_{\mu\nu}(g)\dot{x}^\mu\dot{x}^\nu=0.
\label{eq:geodesic1}
\eeq
Unlike the autoparallel equation \eqref{eq:autoparallel1}, the metric geodesic equation is oblivious to the general affine structure and only cares about the Levi-Civita part entirely determined by the metric, as it should because the length of curves only depends on the metric. We can parameterise the difference between both equations by performing a post-Riemannian expansion $\Gamma^\alpha{}_{\mu\nu}=\bar{\Gamma}^\alpha{}_{\mu\nu}+\Upsilon^\alpha{}_{\mu\nu}$ so the autoparallel equation reads
\beq
\ddot{x}^\alpha+\bar{\Gamma}^\alpha{}_{\mu\nu}(g)\dot{x}^\mu\dot{x}^\nu=-\Upsilon^\alpha{}_{\mu\nu}\dot{x}^\mu\dot{x}^\nu.
\label{eq:autoparallel2}
\eeq
Only experiments can tell us whether particles follow metric geodesic paths or their trajectories are in turn auto-parallel curves for the full affine connection. In other words, we can only constrain the $\Upsilon-$sector by resorting to experiments. However, we can argue which one seems more {\it natural}, with all the caveats that this word might induce, from a theoretical perspective. Let us state our conclusion right away: metric geodesic trajectories seem better aligned with our current understanding of physics. Let us elaborate on why we believe this.

Firstly, the most natural action for a test particle on a gravitational field (that may include a general connection) is given by its line element. If the trajectory of the particle is $x^{\alpha}=x^{\alpha}(\lambda)$ for some affine parameter $\lambda$, we can expect its action to be
\beq
\cS_{\rm pp}=\int g_{\mu\nu}(x)\dot{x}^\mu\dot{x}^\nu\dd\lambda,
\label{eq:ppaction1}
\eeq
which leads to the metric geodesic equation and not to the affine autoparallel one. One might object that the naturalness and our expectation is crucially biased by our prejudice so some more motivation would seem desirable. That \eqref{eq:ppaction1} is the natural action for the gravitational interaction of the particle can be motivated by the fact that the particle's motion should be described by its velocity $\dot{x}^\alpha$ and, in compliance with the equivalence principle, it should reduce to $\eta_{\mu\nu}\dot{x}^\mu \dot{x}^\nu$ in a freely falling frame. Furthermore, once we accept that the particle dof's are described by $\dot{x}^\alpha$, \eqref{eq:ppaction1} can be regarded as the lowest order interaction with the metric tensor from an effective theory perspective. There could be other higher order interactions but they will be suppressed by some appropriate scale. In fact, we do expect higher order corrections of this type. The same reasoning can be applied to determine the coupling to the affine connection. If we stick to the equivalence principle for gravity, then the connection cannot couple directly to the particle. This could be too restrictive because the equivalence principle is only a required consistency coupling prescription for the massless spin-2 sector of the theory \cite{Weinberg:1995mt,Weinberg:1965nx}. However, the connection sector could contain additional propagating dof's that do not need to comply with the equivalence principle so there would not be any reason to impose it for the couplings to the connection. Thus, if we let the connection couple to the particle, the lowest order interaction is given by
\beq
\cS_{\rm pp-\Gamma}=\int \Upsilon_\mu\dd x^\mu=\int \Upsilon_\mu\dot{x}^\mu\dd\lambda,
\label{eq:ppaction2}
\eeq
where $\Upsilon_\mu$ is some arbitrary combination of traces of the connection. The correction to the field equations coming from this coupling is of the form
\beq
\frac{\delta\cS_{\rm pp}}{\delta x^\alpha}\supset \big(\partial_\alpha \Upsilon_\mu-\partial_\mu\Upsilon_\alpha\big)\dot{x}^\mu,
\eeq
which contributes a Lorentz-like force and, certainly, it does not lead to the affine autoparallel equation. Again, we can expect higher order corrections, but they will be suppressed by some suitable scale and it will contain higher powers of the particles velocity. Thus, obtaining the autoparallel equation for the full connection from an appropriate action is substantially more contrived than obtaining the metric geodesic equation, which in turn appears quite naturally. In fact, Eqs. \eqref{eq:autoparallel1} cannot be obtained from a standard variational principle in general. Within the context of teleparallel theories where the curvature vanishes identically, one can design an appropriate variational principle to obtain the corresponding autoparallel equation as suggested in \cite{Fiziev:1995te,Kleinert:1996yi}. One can always resort to suitable constraints and more or less involved couplings leading to the desired equations (whenever this is possible), but this procedure seems artificial to eventually produce the equations in a somewhat ad-hoc manner. An objection to the argument could be that there is no fundamental principle stating that physical equations should follow from an action. After all, not all field equations can be derived from an action principle. Thus, we could regard Eqs. \eqref{eq:autoparallel2} as Lagrange equations of the second kind with some generalised velocity-dependent force precisely given by $\Upsilon^\alpha{}_{\mu\nu}\dot{x}^\mu\dot{x}^\nu$ that go beyond the usual friction forces linear in the velocities and derivable from a Rayleigh dissipation function. However, our current understanding of physics at the most fundamental level can be formulated in terms of the path integral whose primary ingredient is the action (besides an appropriate measure). Let us recall that the standard model of the fundamental interactions including gravity is indeed described by an action so it is natural, though not mandatory, to expect that physical equations should follow from an action principle and, in particular, the motion of particles in a gravito-affine background field.

We will finalise our digression by noticing that a particle is just an idealisation of some more fundamental (classical or quantum) field. Standard bosonic fields like a scalar or spin-1 fields only couple to the metric, so it is difficult to justify the appearance of the connection (other than its Levi-Civita part) in their field equations and, consequently, on the propagation of the associated point-like particles. Furthermore, the propagation of these fields is usually obtained by applying the eikonal or geometric optics approximation to the corresponding hyperbolic equation describing the dynamics of the fields, which in most cases reduces to a wave equation (or a set of them) with the d'Alembertian associated to the metric \cite{Delhom:2020hkb}. In that approximation, the trajectory arises as the curve whose tangent vector is parallel transported with the Levi-Civita connection. On the other hand, we can include couplings to the connection and these will modify the paths of the associated particles in the corresponding approximation, but ensuring that such modifications will lead to the affine autoparallel equation \eqref{eq:autoparallel1} will require a certain amount of artificiality. When considering fermions that do couple to the connection, the conclusion is similar. In that case the eikonal approximation will exhibit additional torsional forces, but they will not mimic the effect of the affine autoparallel propagation \cite{HEHL1971225,Rumpf:1979vh,Audretsch:1981xn,Nomura:1991yx,Cembranos:2018ipn}.

Our discussion here is relevant for concerning the physical importance of {\it geodesically complete} spacetimes in metric-affine theories, meaning spacetimes where the solutions of \eqref{eq:autoparallel1} can be extended to the entire manifold. The incompleteness of these curves can be associated to the existence of singularities. It is then crucial to discern the class of trajectories that carry physical information on the propagation of actual particles. In view of our discussion, it is most natural to consider the solutions of Eqs. \eqref{eq:geodesic1} as the relevant ones in order to draw physical consequences, even if we are in a metric-affine framework. If that is the case and our matter sector couples to the connection directly, then the geodesic equations \eqref{eq:geodesic1} cease to be valid to describe the dynamics of particles because we will need to include the corresponding {\it affine forces}, but these will not, in general, be encapsulated in an autoparallel equation and a case by case study would be required since, as commented above, universality is no longer a property of the interactions. It is also important to emphasise that the metric determining the trajectories of different particles could depend on the species around non-trivial backgrounds, as it is the case for projectively invariant RBG where gravitational waves follow the geodesics of the auxiliary metric $q_{\mu\nu}$, while matter fields travel according to $g_{\mu\nu}$  (see {\it e.g.} \cite{BeltranJimenez:2017doy}).

%%%%%%%%%%%%%%%%%%%%%%%%%%%%%%%
%%%%%%%%%%%%%%%%%%%%%%%%%%%%%%%%%%%%%%%%%%%%%%%

\section{Constrained geometries}\label{sec:GeoConst}

In the precedent sections we have seen that abandoning the projective symmetry in the higher order curvature sector of a metric-affine theory results in the appearance of ghost-like pathologies precisely related to the projective mode. We will now discuss the different frameworks where metric-affine theories can be rendered stable, not by imposing additional symmetries, but by enforcing suitable constraints on the connection, i.e, by restricting to some specific geometries. In this respect, it is known that broad families of theories admit stable (ghost-free) higher order curvature theories for some particular classes of geometries. In this Section we will  review some known examples where the connection is deprived of specific components of the non-metricity and/or torsion. We will finally show a general result that imposing a vanishing torsion reduces general RBG theories to a theory with an extra interacting massive vector field.

\subsection{Torsion-free theories}\label{sec:TorsionFree}

 We will start by showing how imposing a vanishing torsion avoids the presence of ghosts. This general result was already shown in \cite{BeltranJimenez:2019acz}, but we will reproduce here for completeness. The implementation of this constraint can be performed by either only allowing for variations of the symmetric part of the connection (i.e., assuming a symmetric connection from the beginning) or by introducing a set of Lagrange multiplier fields that enforce $\cT^\alpha{}_{\mu\nu}=0$. Either way, the resulting connection equations now read
\beq\label{ConnectionFieldEqsTorsionFree}
\na_\la\lrsq{\sqrt{-q}q^{(\mu\nu)}}-\na_\rho\lrsq{\sqrt{-q}q^{\rho(\mu}}\del^{\nu)}_\la=0.
\eeq
Notice that the only difference with respect to the equations for the unconstrained connection is precisely the trivialisation of their antisymmetric part. Let us decompose $q^{\mu\nu}$ again as in \eqref{metricsplitting}. Due to the vanishing of the torsion tensor, the general decomposition of the connection \eqref{conndecomp} lacks the contortion tensor. Thus, the connection can here be split in a Levi-Civita connection of $h^{\mu\nu}$ and a disformation part that depends on the non-metricity $N_{\lambda\mu\nu}\equiv \nabla^h_\lambda h_{\mu\nu}$ as\footnote{This splitting allows us to write a general affine connection in terms of the torsion, an arbitrary invertible symmetric 2-tensor, its first derivatives and its covariant derivative (i.e. its non-metricity with respect to $\Ga$).}
\beq\label{ConnectionDeccomposition}
\Ga_{\mu\nu}^\al=\bar{\Ga}_{\mu\nu}^\al(h)+L^\al{}_{\mu\nu}(N)
\eeq
without loss of generality, where the disformation tensor is now built with the non-metricity of $h^{\mu\nu}$. The above splitting allows to obtain the following relations that we will use below
\begin{align}
\nabla_\lambda\big(\sqrt{-h}h^{\lambda\nu}\big)&=\sqrt{-h}\bL^\nu,\label{L1}\\
\nabla_\lambda\big(\sqrt{-h}B^{\lambda\nu}\big)&=\sqrt{-h}\nabla^h_\lambda B^{\lambda\nu},\label{L2}
\end{align}
where $\bL^\nu\equiv L^\nu{}_{\alpha\beta} h^{\alpha\beta}$ is one of the two independent traces of the disformation tensor. The trace of the connection equation (\ref{ConnectionFieldEqsTorsionFree}) together with \eqref{L1} yields 
\beq
\nabla^h_\lambda B^{\lambda \nu}=\frac{1-D}{1+D}\bL^\nu,
\label{eq:constraint1}
\eeq
which implies the dynamical constraint
\beq
\nabla^h_\nu \bL^\nu=0.
\label{eq:constraint2}
\eeq
On the other hand, contracting the connection equation (\ref{ConnectionFieldEqsTorsionFree}), with $h_{\mu\nu}$ defined as the inverse of $h^{\mu\nu}$, leads to 
\beq
L_\mu=\frac{2}{(2-D)(1+D)}\bL_\mu,
\eeq
where $L_{\mu}\equiv L^\alpha{}_{\mu\alpha}$ and indices are raised and lowered with $h_{\mu\nu}$. Thus, we see that there is only one independent trace of the disformation tensor. Using the above relations in the connection equation \eqref{ConnectionFieldEqsTorsionFree}, we are led to
\beq
2h^{\alpha(\mu} L^{\nu)}{}_{\lambda\alpha}=L_\lambda h^{\mu\nu}+(2-D) L_\alpha h^{\alpha(\mu}\delta^{\nu)}{}_\lambda.
\label{eq:connectionTfree2}
\eeq
Given that the the non-metricity tensor of the auxiliary metric
is given by $N_\lambda{}^{\mu\nu}\equiv -\nabla_\lambda h^{\mu\nu}=-2h^{\alpha(\mu}L^{\nu)}{}_{\lambda\alpha}$, which implies the identity $L_\mu=-\frac12h_{\alpha\beta}N_\mu{}^{\alpha\beta}\equiv-\frac12 \tilde{N}_\mu$, the above equation can be used to re-write the connection equation (\ref{eq:connectionTfree2}) as a constraint for the non-metricity tensor
\beq\label{relationtracesL}
N_\lambda{}^{\mu\nu}=\frac12\Big[ \tilde{N}_\lambda h^{\mu\nu}+(2-D)\tilde{N}_\alpha h^{\alpha(\mu}\delta^{\nu)}_\lambda\Big],
\eeq
which becomes completely specified by its Weyl component (although it is not Weyl-like). Thus we see that the connection field equations can be fully solved explicitly, and the connection is given by a disformation piece given by the non-metricity tensor \eqref{relationtracesL} added to the Levi-Civita of $h^{\mu\nu}$. Given that this disformaton piece is completely determined by $\tilde{N}_\mu$ (the Weyl trace of the non-metricity of $h^{\mu\nu}$) , the connection carries only one additional vector component, instead of a vector field plus a 2-form as in the most general case. Moreover, from  the transversality constraint (\ref{eq:constraint2}) obtained above, this new vectorial component must be a Proca field, thus propagating only three extra degrees of freedom. The corresponding metric equations of the system will allow to solve algebraically for $h^{\mu\nu}$ as a function of the matter fields and (possibly) the new vector field $\tilde{N}_\mu$, which ensures the absence of the pathologies that were found in the most general case. To illustrate this, let us re-consider a particular example that has already been treated in the literature. Assume a metric-affine gravitational Lagrangian of the form 
\begin{equation}
 \mathcal{L}=\cR+c_1 \cR_{[\mu\nu]}\cR^{[\mu\nu]}.   
\end{equation} 
As explained above, this theory breaks projective symmetry due to the presence of the antisymmetric part of the Ricci in the action. Therefore pathologies should arise in the general case unless further constraints are imposed. However, as shown in past works \cite{Buchdahl:1979ut,Vitagliano:2010pq},  the torsion-free version of this model reduces to the Einstein-Proca system , where the Proca field arises from the connection sector. For more general examples with violation of projective symmetry but where the torsion-free constraint is imposed torsion, the Proca field will in general develop non-trivial interactions, as was already discussed in  \cite{Olmo:2013lta} for the Ricci-based sub-family $F(g^{\mu\nu}, \cR^{\mu\nu}\cR_{\mu\nu})$ with the torsion-free constraint.\\

To enlighten the mechanism that renders the torsion-free version of generalised RBG theories ghost-free, let us resort to the the Einstein frame of the theory making explicit the torsion-free constraint. The action of the theory can be written as
\begin{align}\label{NoTorsionEinstein}
\cS=&\frac12\int\dd^Dx\sqrt{-g}\Big[f(\Sigma,A)+\frac{\partial f}{\partial \Sigma_{\mu\nu}}\big(\cR_{(\mu\nu)}-\Sigma_{\mu\nu}\big)+\frac{\partial f}{\partial A_{\mu\nu}}\big(\cR_{[\mu\nu]}-A_{\mu\nu}\big)+\frac{1}{\sqrt{-g}}\lambda_\alpha{}^{\mu\nu}\cT^\alpha{}_{\mu\nu}\Big],
\end{align}
where $\lambda_\alpha{}^{\mu\nu}$ is a Lagrange multiplier that enforces the torsion-free constraint $\cT^\alpha{}_{\mu\nu}=0$; and $A_{\mu\nu}$ and $\Sigma_{\mu\nu}$ are auxiliary fields that are antisymmetric and symmetric respectively. In an analogue manner to sec. \ref{sec:NSGFrame}, we can perform field re-definitions which allow us to algebraically solve for the space-time metric $g^{\mu\nu}$ in terms of $h^{\mu\nu}$, $B^{\mu\nu}$ and the matter fields; thus integrating $g^{\mu\nu}$ out. We can then write the Einstein frame action for torsion-free generalised RBGs as
\begin{align}
\cS=&\frac12\int\dd^Dx\Big[\sqrt{-h}h^{\mu\nu}\cR_{(\mu\nu)}+\sqrt{-h}B^{\mu\nu}\cR_{[\mu\nu]}+\cU(h,B,T)+\lambda_\alpha{}^{\mu\nu}\cT^\alpha{}_{\mu\nu}\Big].
\end{align}
This action gives the same connection equations that we solved above \eqref{ConnectionFieldEqsTorsionFree}, so we can take the above solution (basically the splitting (\ref{ConnectionDeccomposition}) and equation (\ref{relationtracesL}) together)  and plug it back into the above action. As it can be seen, the solution for the connection satisfies the relations 
\begin{align}
\cR_{[\mu\nu]}=&-\frac{1}{2}\partial_{[\mu}\tilde{N}_{\nu]},\nonumber \\
\cR_{(\mu\nu)}=&R_{\mu\nu}(h)+\frac{(D-2)(D-1)}{16}\tilde{N}_\mu \tilde{N}_\nu-\frac{(D-1)}{4}h_{\mu\nu}\nabla^h_\alpha \tilde{N}^\alpha
\end{align}
which, after dropping the surface term $\nabla^h_\mu \tilde{N}^\mu$, allow us to re-express the action \eqref{NoTorsionEinstein} in terms of the metric $h_{\mu\nu}$, the 2-form $B_{\mu\nu}$ and the vector field $\tilde{N}^\mu$ as
\begin{align}
\begin{split}
\cS=\frac12\int\dd^Dx\Big[\sqrt{-h}\Big(&R(h)+\frac{(D-2)(D-1)}{16}\tilde{N}^2-\frac{1}{2}B^{\mu\nu}\partial_{[\mu}\tilde{N}_{\nu]}\Big)+\cU(h,B,T)\Big],
\end{split}
\label{eq:finalaction}
\end{align}
Notice that this form of the action reproduces the constraint on the 2-form (\ref{eq:constraint1}) as the field equations of the vector field $\tilde{N}^\mu$ (which are in some sense the connection equations in the corresponding RBG frame), which read
\beq
\nabla^h_\mu B^{\mu\nu}=-\frac{(D-2)(D-1)}{4}\tilde{N}^\nu,
\eeq
and imply the constraint $\nabla^h_\alpha \tilde{N}^\alpha=0$. At the same time the 2-form field equations yield a non-linear relation among the 2-form, the field-strength of the vector field $\tilde{N}^\mu$, and the matter fields given by 
\beq\label{ProcaEqPotential}
\partial_{[\mu}\tilde{N}_{\nu]}=\frac{2}{\sqrt{-h}}\frac{\partial{\cU}}{\partial B^{\mu\nu}}.
\eeq
 This stems from the fact that our final action (\ref{eq:finalaction}) is nothing but the first-order form of a self-interacting massive vector field coupled to the matter sector. Going back to the particular case $F=\cR+c_1 \cR_{[\mu\nu]}\cR^{[\mu\nu]}$, we can reproduce the above results (found previously in \cite{Buchdahl:1979ut,Vitagliano:2010pq,Olmo:2013lta}). For this particular example, the metric $h^{\mu\nu}$ is exactly $g^{\mu\nu}$, the 2-form is given by $B^{\mu\nu}=2c_1\cR^{[\mu\nu]}$, and the effective potential reads $\cU=-(\sqrt{-h}/4c_1) B^{\mu\nu}B_{\mu\nu}$. Thus \eqref{ProcaEqPotential} becomes
 \beq
 \text{d}\tilde{N}=\mathcal{F}
 \eeq
showing that (\ref{eq:finalaction}) is a first-order description of a free Proca field $\tilde{N}_\mu$ with field-strength given by $\mathcal{F}_{\mu\nu}=(2/c_1)B_{\mu\nu}$.

\subsection{Weyl geometries}
Let us now briefly comment on another paradigmatic extension of the Riemannian framework introduced by Weyl shortly after the GR inception which has been analised widely in the literature (see e.g. the nice survey in \cite{Scholz:2011za}). This geometry is characterised by local scale (gauge) invariance and a torsion-free connection so the only non-trivial part of the affine connection is the so-called Weyl trace of the non-metricity $A_\alpha=-\frac{2}{D}g^{\mu\nu}Q_{\alpha\mu\nu}$. This allows to replace the metric compatibility condition $\bar\nabla_\alpha g_{\mu\nu}=0$ by $D_\alpha g_{\mu\nu}\equiv(\bar{\nabla}_\alpha-A_\alpha)g_{\mu\nu}=0$ which is invariant under the scale transformation
 $g_{\alpha\beta}\rightarrow e^{2\alpha(x)} g_{\alpha\beta}$, under which $A_\mu$ transforms as $A_\mu\rightarrow A_\mu-\partial_\mu \alpha$ as required by invariance of the affine connection.

Theories whose actions are constructed in terms of quadratic curvature invariants for a Weyl connection trivially admit ghost-free theories and, consequently, imposing the connection to be of the Weyl form evidently avoids the ghostly pathologies of the general RBG theories. This constraint can be implemented either by imposing the connection to be Weyl-like from the beginning or by adding suitable Lagrange multipliers. Now we should impose a vanishing torsion and also vanishing of all the non-metricity irreducible components except for the Weyl trace. Since for the torsion-free case there are no ghostly degrees of freedom, it is clear that for Weyl geometries, since they are a sub-class of the torsion-free ones, which also feature additional constraints (non-metricity is forced to be vectorial), there will be no ghosts either. General quadratic theories in Weyl geometries have been studied in e.g. \cite{Jimenez:2014rna} where it was shown that some interesting non-trivial interactions for the Weyl vector can be generated.

\subsection{Geometries with vector distortion}\label{sec:Vecdir}

The affine connection in Weyl geometries are characterised by a vector field that controls the departure from the Levi Civita connection. A natural generalisation is to include not only this vector part, but a general vector piece of the connection in both the torsion and the non-metricity. Such a general connection was considered in \cite{Aringazin_1991} in the absence of torsion and was extended to include the torsion trace in \cite{Jimenez:2015fva,Jimenez:2016opp}. The connection in these geometries can be parameterised as
\beq
\Gamma^\alpha{}_{\mu\nu}=\bar{\Gamma}^\alpha{}_{\mu\nu}-b_1A^\alpha g_{\beta\gamma}+b_2\delta^\alpha_{(\beta} A_{\gamma)}+b_3\delta^\alpha_{[\beta} A_{\gamma]}+b_4\epsilon^\alpha{}_{\mu\nu\rho} S^\rho.
\eeq
This is the minimal field content to describe the desired geometrical setup. It is necessary to have at least two different vector fields with opposite transformation properties under parity in order to account for the axial part of the torsion. The remaining vector pieces, i.e., the two non-metricity traces and the torsion trace, have been identified (up to some proportionality constant) so that this sector is fully described by one single vector field. It would be interesting to study the geometries where the different vector pieces are not identified and the presence of some internal symmetries in that sector (see \cite{Iosifidis:2019fsh} related to this point). The present framework however allows to substantially simplify the analysis. Within the framework of curvature-based theories, the general quadratic action can be written as
 \begin{align}
\cS_{\rm VD}  = &M_{\rm 2} \int \dd^D x \sqrt{-g}\Big[\cR^2 + \cR_{\alpha\beta\gamma\delta}\Big( d_1 \cR^{\alpha\beta\gamma\delta} + d_2 \cR^{\gamma\delta\alpha\beta} 
-  d_3 \cR^{\alpha\beta\delta\gamma}\Big)\\
& -  4\Big( c_1 \cR_{\mu\nu}\cR^{\mu\nu} +  c_2 \cR_{\mu\nu}\cR^{\nu\mu} 
+  \cP_{\mu\nu}\left( c_3 \cP^{\mu\nu} + c_4 \cP^{\nu\mu} -  c_5 \cR^{\mu\nu} - c_6 \cR^{\nu\mu}\right) 
+  \cQ_{\mu\nu}(c_7 \cQ^{\mu\nu} + c_8 \cR^{\mu\nu}+ c_9\cP^{\mu\nu})\Big)\,   \nonumber
\Big]\,.
\label{SquadraticVD}
\end{align}
where $d_i$ and $b_i$ are some dimensionless constants and $M_2$ some scale. This action will generically lead to instabilities, once again along the lines of what one would expect as discussed in detail above. In order to guarantee a ghost-free pure graviton sector, it is convenient to impose that the theory reduces to a Gauss-Bonnet theory in the Riemannian limit, i.e., when $A_\mu\rightarrow 0$. It is then remarkable that it is sufficient to restrict the geometrical framework rather than the parameters in the action in order to obtain a ghost-free vector-tensor theory \cite{Jimenez:2015fva,Jimenez:2016opp}. The ghost-free geometries are characterised by $2b_1-b_2-b_3=0$ and the resulting action reduces to
\begin{align}
\cS_{\rm VD}=\mu\int\dd^Dx\sqrt{-g}&\left[\Big(R^2-4R_{\mu\nu}R^{\mu\nu}+R_{\mu\nu\rho\sigma}R^{\mu\nu\rho\sigma}\Big)-\frac\alpha 4 F_{\mu\nu} F^{\mu\nu}+\xi A^2\nabla\cdot A -\lambda A^4-\beta G^{\mu\nu} A_\mu A_\nu\right]
\end{align}
where $\alpha$, $\xi$, $\lambda$ and $\beta$ are some constants that are given in terms of the parameters in \eqref{SquadraticVD} and  $F_{\mu\nu}=\partial_\mu A_\nu -\partial_\nu A_\mu$. The noteworthy property of this action is that the vector field features derivative non-gauge invariant interactions and a non-minimal coupling, but which precisely belong to the class of ghost-free interactions {\color{red}cite}. Thus, the general result regarding the ghostly pathologies has been resolved in the vector distorted geometries by two conditions, namely: $i)$ imposing the recovery of the safe Gauss-Bonnet quadratic gravity in the absence of distortion and $ii)$ restricting the class of geometries. The singular property of the selected ghost-free geometries is that they generalise the Weyl connection by including a trace torsion piece but maintaining the Weyl invariance of the metric (in)-compatibility condition. This can be easily understood by noticing that the non-metricity for this restricted class of geometries is $Q_{\mu\alpha\beta}=(b_3-b_2) A_\mu g_{\alpha\beta}$ which is of the Weyl type. However the torsion is non-vanishing and given by $\cT^\alpha{}_{\mu\nu}=2b_3\delta^\alpha_{[\mu}A_{\nu]}$. We refer to \cite{Jimenez:2015fva,Jimenez:2016opp} for the detail discussion on the interesting geometrical properties of these geometries and here we will content ourselves with simply signalling how ghost-free theories can be obtained.

%%%%%%%%%%%%%%%%%%%%%%%%%%%%%%%%%%%%%%%%%%%

%%%%%%%%%%%%%%%%%%%%%%%%%%%%%%%%%%%%%%%%%%

\subsection{Riemann-Cartan geometries}\label{sec:NMfree}
Let us now consider the case of one of the first extensions of GR, namely the extension of the Riemannian framework to the so-called Riemann-Cartan geometry, where the connection is allowed to have a torsion component while keeping a trivial non-metricity. This can be achieved by introducing a suitable Lagrange multiplier in the action \eqref{GeneralAction}:
\beq\label{GeneralActionwithNMmult}
\cS[g_{\mu\nu},\Ga,\lambda]=\frac12\int \dd^Dx \sqrt{-g}\,\lrsq{F\big(g^{\mu\nu},\cR_{\mu\nu}(\Ga)\big)+\lambda^{\alpha}{}_{\mu\nu}\na_\al g^{\mu\nu}}+\cS_{\rm m}[g_{\mu\nu},\Psi].
\eeq
While the torsion-free constraint heals the instabilities of generalised RBGs, this is not the case for a constraint imposing the vanishing of the  non-metricity tensor. Given that the full analysis is rather cumbersome in this case, we will simply highlight the main differences between the vanishing non-metricity and vanishing torsion constraints, emphasising which are the conditions that improve the pathological behaviour of generalised RBGs in their torsion-free versions that do not occur when the non-metricity free constraint is imposed. First of all notice that varying the above action with respect to $\lambda^\al{}_{\mu\nu}$ one gets the constraint $\na_\al g^{\mu\nu}=-Q_\al{}^{\mu\nu}=0$. Now an infinitesimal variation of the above action with respect to the connection yields
\beq
\begin{split}\label{VarActionNMfree}
\delta_\Ga\cS=&\frac12\int \dd^Dx \sqrt{-g}\frac{\partial F}{\partial\cR_{\mu\nu}}\delta_\Ga \cR_{\mu\nu}=-\frac12\int \dd^Dx \sqrt{-q}q^{\mu\nu}\lr{\na_\al\delta\Ga^\al{}_{\nu\mu}-\na_\nu\delta\Ga^\al{}_{\al\mu}-\cT^\lambda{}_{\nu\al}\delta\Ga^\al{}_{\la\mu}}
\end{split}
\eeq
where the conditions $Q^\al{}_{\mu\nu}=0$ and $\delta_\Ga Q_{\al}{}^{\mu\nu}=0\rightarrow \delta_\Ga L^{\al}{}_{\mu\nu}=0$ are imposed by the Lagrange multiplier field equation after integrating it out. The root of the difference between the two cases is the term in the variation of the Ricci tensor \eqref{transfRicci}. In the above variation of the action, that term vanishes in the torsion-free case (after integrating out the vanishing torsion field), while this does not occur in the non-metricity case. As a consequence, the connection field equations for the vanishing non-metricity case are
\begin{eqnarray}
\na_\la \lrsq{\sqrt{-q}q^{\nu\mu}}-\del^\mu{}_\la\na_\rho\lrsq{ \sqrt{-q}q^{\nu\rho}}
&=&\sqrt{-q}\lrsq{\cT^\mu{}_{\la\al} q^{\nu\al}+\cT^\al{}_{\al\la} q^{\nu\mu}-\del^\mu{}_\la\cT^\al{}_{\al\be} q^{\nu\be}},
\end{eqnarray}
thus having the same tensorial structure than the ones in the general case\footnote{Notice that here we could drop the $\sqrt{-q}$ from the connection field equations by defining $q^{\mu\nu}\equiv\partial F/\cR_{\mu\nu}$. However since it does not introduce any advantage, we will not do it to facilitate the comparison with the torsion-free case.} \eqref{VariationConnection}, which does not happen in the torsion-free case \eqref{ConnectionFieldEqsTorsionFree}. This difference will have consequences in the number of degrees of freedom propagated in the different cases, as well as in their stability properties. To make this more clear, let us first decompose the non-metricity free connection as $\Gamma^\al{}_{\mu\nu}=\bar\Gamma^\al{}_{\mu\nu}(q)+L^\al{}_{\mu\nu}+K^\al{}_{\mu\nu}$. Notice that although the covariant derivative $\na_\al g_{\mu\nu}$ vanishes, this is not true for $\na_\al h_{\mu\nu}$, and thus the distortion tensor corresponding to $h_{\mu\nu}$ in the connection decomposition is non-vanishing. We thus see that the non-metricity free condition does not have an implementation as nice as the torsion-free condition, and the structure of the equations is identical to the general case, having also the constraint
\begin{equation}\label{ConstraintBNM}
\na^h_\la B^{\la\mu}=0.
\end{equation}
 In the torsion-free case, we found instead that the divergence (with respect to $h^{\mu\nu}$) of the 2-form was proportional to one of the traces of the distortion tensor $\bL_\mu$. Thus, in both the torsion-free and the non-metricity-free cases the divergence of the 2-form can be eliminated from the field equations. Another important point is that the absence of $K^\al{}_{\mu\nu}$ in the torsion-free case and the index symmetries of $B^{\mu\nu}$ and $L^\al{}_{\mu\nu}$ yield the relations\eqref{L1} and \eqref{L2}. While \eqref{L1} is still occurring in this case, the analogue relation to \eqref{L2} is 
\begin{align}
\nabla_\lambda (\sqrt{-h}B^{\lambda\nu})=\sqrt{-h}\nabla^h_\lambda B^{\lambda\nu}+\sqrt{-h}\lr{t_\al B^{\al\nu}+\frac{1}{2}\cT^{\nu}{}_{\al\be}B^{\al\be}},\label{L2NM}
\end{align}
where $t_\al\equiv\cT^\be{}_{\be\al}$ and the first term on the right hand side vanishes due to \eqref{ConstraintBNM}. Thus, while in the torsion-free case these relations together with the divergence of the two-form \eqref{eq:constraint1} allow to write $\na_\al(\sqrt{-h}h^{\al\mu})$ and $\na_\al(\sqrt{-h}B^{\al\mu})$ in terms of the vector field $\bL_\mu$, this is not the case in the non-metricity free scenario. Once we have highlighted these differences, which rely only on the decomposition of the connection that one is able to do in the different cases, we are now ready to understand why the difference in the tensorial structure of the connection field equations plays a crucial role in the stability properties. After the decomposition of $q^{\mu\nu}$ into its symmetric and antisymmetric parts, due to the symmetrization of $\mu$ and $\nu$ in the torsion-free case only the contraction $\na_\al B^{\al\mu}$ enters the connection field equations. As explained above, this can be substituted by $\bL_\mu$ in the torsion-free case and, together with the relations \eqref{L1} and \eqref{L2}, it allows to find a relation between both traces of the distortion tensor. Then, since $\na_\al  B^{\mu\nu}$ does not appear in the equations, and $\na_\al  h^{\mu\nu}$ can be written only in terms of $L^\al{}_{\mu\nu}$, the connection equation \eqref{ConnectionFieldEqsTorsionFree} allows to find a solution for the full connection as the Levi-Civita conection of the auxiliary metric plus a distortion part characterized only by the vector field $\bL_\mu$. In contrast, since in the vanishing non-metricity case (as in the general one) the symmetrization of $\mu$ and $\nu$ does not occur in the connection field equations, not only its trace but also the full covariant derivative of $B^{\mu\nu}$ enters the connection field equations. This makes $B_{\mu\nu}$ a propagating field and makes it impossible to solve the connection only in terms of a new vector field. Indeed it can be seen that the torsion tensor has the schematic form $\na B/(1+B)$ as happened to $\Upsilonh$ in sec. \ref{sec:GenRBG}, thus potentially introducing Ostrogradski instabilities propagated by the 2-form. Therefore the Einstein frame version of this theory would be formally identical to the one of the general case, since the distortion of $h^{\mu\nu}$ is not vanishing here. Thus we see that in general the constraint of vanishing non-metricity will not heal the instabilities of the previous theory, as the extra 5 degrees of freedom corresponding to the projective mode an the 2-form will in general also propagate, although there could be fine-tuned Lagrangians in which this does not occur.

We will end this Section by noticing that the Poincar\'e gauge theories \cite{Kibble:1961ba} are formulated in a Riemann-Cartan geometry. It is known that the general quadratic theories of this class present pathologies and only very specific choices of parameters give rise to healthy theories (see e.g. \cite{Hayashi:1979wj,Sezgin:1979zf,Yo:1999ex,Blagojevic:2018dpz,Vasilev:2017twr,Jimenez:2019tkx,Jimenez:2019qjc}). As repeated several times, it is possible to have phenomenologically viable theories by interpreting them as effective field theories as done in \cite{Aoki:2019snr,Aoki:2020zqm}.

%%%%%%%%%%%%%%%%%%%%%%%%%%%%%%%%%%%%%%%%%%
%%%%%%%%%%%%%%%%%%%%%%%%%%%%%%%%%%%%%%%%%%

\section{Hybrid theories}\label{sec:Hybrid}
So far we have considered RBG in the pure metric-affine formalism so that only the curvature of the full connection enters the action. As we explained in Sec. \ref{sec:Generalities}, every spacetime endowed with a metric tensor admits a distinguished connection given by the Christoffel symbols of the metric. Thus, for any spacetime with a general connection there is a coexistent affine structure provided by the Levi-Civita connection. The hybrid formalism \cite{Harko:2011nh, Capozziello:2015lza} steps outside the purely metric-affine framework and embraces these two coexisting affine structures so that the action contains the curvatures of the two connections. As we will see, rather than improving the situation of the pure metric-affine formalism, delving into the hybrid framework generically introduces even more pathologies. This may not be too surprising since the hybrid formalism is prone to the independent pathologies of the metric and metric-affine formalisms separately from the outset and hence it is natural to expect the same pathologies at the very least. The existence of pathologies in the hybrid formalism was analysed in \cite{Koivisto:2013kwa} by looking at the propagator on flat spacetime and identifying the presence of ghosts for a class of hybrid theories whose action is an arbitrary function of the two Ricci scalars $R(g)$ and $\cR(\gamma)$ and the hybrid Ricci square term $R_{\mu\nu}(g)\cR^{\mu\nu}(\Gamma)$.

In order to pinpoint the sources of pathologies for the hybrid theories, we will consider the following hybrid action
\beq
\cS_{\rm hybrid}=\int\dd^Dx\sqrt{-g}f(\cR_{\mu\nu},R_{\mu\nu}).
\eeq
We will then proceed analogously to the pure metric-affine formalism to write the action as
\beq
\cS_{\rm hybrid}=\int\dd^Dx\Big[\sqrt{-q} q^{\mu\nu}\cR_{\mu\nu}(\Gamma)+\cU(R_{\mu\nu},q,g)\Big]
\label{Eq:hybrid2}
\eeq
where we have defined
\beq
\cU\equiv\sqrt{-g}\left[f-\frac{\partial f}{\partial\Sigma_{\mu\nu}} \Sigma_{\mu\nu}\right],\quad\text{and}\quad \sqrt{-q}q^{\mu\nu}\equiv \sqrt{-g}\frac{\partial f}{\partial\Sigma_{\mu\nu}},
\label{eq:defUq}
\eeq
and here $f$ is understood as a function of $\Sigma_{\mu\nu}$ and $R_{\mu\nu}$. The general hybrid action written in the form \eqref{Eq:hybrid2} is sufficient to understand the multiple sources of instabilities. Since we have linearised in the Ricci of the connection, that sector alone already reproduces the pathologies associated to the projective mode and the additional 2-form field that we have extensively discussed in precedent sections. Furthermore, even if we impose a projective symmetry in an attempt to avoid those pathologies, we can then straightforwardly integrate out the connection and obtain the equivalent action
\beq
\cS_{\rm hybrid}=\int\dd^Dx\Big[\sqrt{-q} q^{\mu\nu}\cR_{(\mu\nu)}(q)+\cU(R_{\mu\nu},q,g)\Big],
\label{Eq:hybrid3}
\eeq
so we have an Einstein-Hilbert term to describe the dynamics of the (now symmetric) field $q_{\mu\nu}$. That pure metric-affine sector is then fine. However, the hybrid couplings introduce yet two additional sources of pathologies. On one hand, if we have an arbitrary dependence on the metric Ricci tensor, the theory will be prone to the usual Ostrogradski instabilities in the metric sector. Furthermore, even if we avoid those problems by utilising only the Ricci scalar of the metric for instance, that is known to represent a safe higher order curvature of the metric formalism, the potential $\cU$ will introduce arbitrary interactions between $q_{\mu\nu}$ and $g_{\mu\nu}$ so we will have an interacting bi-metric theory that will again introduce ghostly modes unless much care is taken in the construction of the interactions. We can understand this a bit better by considering a simplified theory where the metric and metric-affine sectors are split as
\beq
\cS_{\rm hybrid}=\int\dd^Dx\sqrt{-g}\left[\frac12 R(g)+\mathcal{F}(\cR_{(\mu\nu)})\right]
\eeq
where we have separated the pure metric sector described by the Einstein-Hilbert action and the metric-affine sector on which we have imposed a projective symmetry. Each of these sectors by itself would seem perfectly fine. However, they can talk to each other through the $\sqrt{-g}$ factor in the volume element and this will be the source of the problems. In view of our results above and neglecting matter fields for simplicity, we can expect to have two Einstein-Hilbert terms once we integrate the connection out. This is in fact the case, but we also generate a potential so the action reads
\beq
\cS_{\rm hybrid}=\int\dd^Dx\left[\frac{\sqrt{-q}}{2}q^{\mu\nu}\cR_{\mu\nu}(\Gamma)+\frac{\sqrt{-g}}{2} R(g)+\mathcal{U}(q,g)\right],
\label{Eq:hybridbi-metric3}
\eeq
where the dependence on the general potential term in \eqref{Eq:hybrid2} can be separated as the $R(g)$ term in the above action. The resulting action is then a bi-metric theory where the two metrics interact through the potential $\cU$ and it will suffer from a Boulware-Deser ghost \cite{Boulware:1973my}. Since this potential is determined by the function $f$, only functions that generate the known ghost-free potentials \cite{deRham:2010kj,Hassan:2011zd} have a chance to be stable. It is clear that resorting to a hybrid action not only cannot cure the found instabilities in RBG theories, but makes things even worse by introducing yet new sources of ghosts. A way around this general no-go result for stable hybrid theories results in theories where the bi-metric construction fails. This happens for theories where only the Ricci scalars are allowed, i.e., theories described by the action
\beq
\cS_{\rm hybrid}=\int\dd^Dx\sqrt{-g}f(\cR,R).
\eeq
We can proceed analogously by performing the corresponding Legendre transformations to linearise in $R$ and $\cR$, but now we only need to introduce two auxiliary scalar fields instead of the tensor $\Sigma_{\mu\nu}$ so we can rewrite the action as
\beq
\cS_{\rm hybrid}=\int\dd^Dx\Big[\sqrt{-g}\,\chi\, g^{\mu\nu}\cR_{\mu\nu}(\Gamma)+\sqrt{-g}\varphi g^{\mu\nu} R_{\mu\nu}+\cU(\varphi,\chi)\Big].
\eeq
From this action we see that now the connection is nothing but the Levi-Civita connection of a metric that is conformally related to $g_{\mu\nu}$. In order words, the definition of $q^{\mu\nu}$ in \eqref{eq:defUq} yields $q_{\mu\nu}=\tchi g_{\mu\nu}$, with $\tchi=\chi^{\frac{2}{D-2}}$, so we only introduce an extra scalar instead of the full symmetric $q_{\mu\nu}$. The action then takes the form
\beq
\cS_{\rm hybrid}=\int\dd^Dx\sqrt{-g}\Big[(\varphi+\chi)R+2(1-D)\chi\Big(\Box\log\tchi+\frac{D^2-4D-4}{2}(\partial\log\tchi)^2\Big)+\cU(\varphi,\chi)\Big].
\eeq
It is then apparent that these theories propagate two additional scalars and avoid the Boulware-Deser ghosts of the general case. It was found in \cite{Koivisto:2013kwa} however, that even these theories seem to present some tension between the absence of tachyons and ghosts around a flat Minkowski background so it is unavoidable to have some kind of instabilities.

\section{Conclusions}

In this work we have addressed the unstable nature of general gravitational theories in a metric-affine formalism constructed with arbitrary curvature invariants. In particular, we showed in section \ref{sec:GenRBG} the crucial role player by the projective symmetry in RBGs and how its breaking generically leads to pathologies. Remarkably, we could establish an interesting relation between RBGs without projective symmetry and non-symmetric gravity that allows to relate the found instabilities in both families of theories and, furthermore, give a novel interpretation of the pathologies in non-symmetric gravity. We have traced the origin of the pathologies to the absence of a proper kinetic term for the projective mode as well as the presence of non-minimal couplings the 2-form field associated to the antisymmetric part of the metric. This is shown to happen in two independent forms: in section \ref{sec:InstDecLim} by taking the decoupling limit after applying the S\"teckelberg trick to restore the gauge symmetry for the 2-form, and in section \ref{sec:ProblemAdditionalDOFs} by explicitly writing the action in a pure post-Riemannian form with two extra dynamical fields corresponding to the 2-form and the projective mode. Moreover, in section \ref{sec:SolConn} we also present a formal solution that can be used to perturbatively solve the connection which makes explicit that the 2-form features unstable couplings as was the case for NGT \cite{Damour:1992bt}. The possibility that some couplings between matter and the connection could cure these instabilities is also explored in section \ref{sec:Mattercoupling}. First we sketch the construction of the non-symmetric gravity frame for generalized RBGs in presence of couplings between matter and connection in \ref{sec:NGTframemat}, arguing that the addition of such couplings will generally be oblivious to the ghosts (if it does not worsen the instabilities). We also elaborate in section \ref{sec:geodesics} on the distinction between metric geodesics and autoparallel curves and discuss why metric geodesics seem more natural trajectories for freely falling particles than affine ones.

Motivated by the results of \cite{BeltranJimenez:2019acz} that constraining torsion to vanish can render generalized RBGs stable even with broken projective symmetry, we have extended the discussion to more general geometrical constraints in section \ref{sec:GeoConst}, surveying the cases of torsionless spacetimes, Weyl geometries, geometries with vector distorsion and Riemann-Cartan geometries. In all cases, although pathologies can still be present, we have explicitly shown particular cases where the theories are ghost-free. Finally, we have shown how theories formulated in the hybrid formalism are also generically prone to ghost-like pathologies (in particular they contain a Boulaware-Deser ghost) even in a more harmful manner.

In \ref{sec:OnMoreGeneral} we have argued why these theories are generically plagued by ghost-like instabilities and the results presented in this work should make it clear. Thus, although there will certainly be healthy metric-affine theories with higher order curvature invariants, it is crucial to guarantee the avoidance of pathologies before reliably using them for any physical application in e.g. cosmology or black hole physics.

%%%%%%%%%%%%%%%%%%%%%%%%%%%
\acknowledgments 
We thank useful discussions with Tomi S. Koivisto, Gonzalo J. Olmo, Damianos Iosifidis and Alejandro Jim\'enez-Cano.
JBJ acknowledges support from the  {\textit{ Atracci\'on del Talento Cient\'ifico en Salamanca}} programme and the MINECO's projects FIS2014-52837-P and FIS2016-78859-P (AEI/FEDER). A.D. is supported by  an FPU fellowship. This work is supported by the the Spanish Projects No. FIS2017-84440-C2-1-P (MINECO/FEDER, EU), the Project No. H2020-MSCA-RISE-2017 Grant No. FunFiCO-777740, Project No. SEJI/2017/042 (Generalitat Valenciana), the Consolider Program CPANPHY-1205388, and the Severo Ochoa Grant No. SEV-2014-0398 (Spain). This article is based upon work from COST Action CA15117, supported by COST (European Cooperation in Science and Technology). AD also wants to thank hospitality to the \textit{Departamento de F\'isica Te\'orica de la Universidad de Salamanca}.
%%%%%%%%%%%%%%%%%%%%%%%%%%%

\appendix
\section{Degrees of freedom of massive and massless 2-forms}\label{sec:Appendix2-form}
Let us consider the action for a massless 2-form in $D=4$:
\beq
\cS=-\frac{1}{12}\int\dd^4x\sqrt{-g} H^2
\eeq
with $H_{\al\be\ga}\equiv\partial_{[\al}B_{\be\ga]}$. The number of degrees of freedom contained in the 2-form can be readily identified by dualising this action. To that end, let us rewrite it in its first order form
\beq
\cS=-\frac16\int\dd^4x\sqrt{-g}\left(\Pi^{\alpha\beta\gamma}\partial_{[\alpha} B_{\beta\gamma]}-\frac12\Pi^2\right),
\eeq
where $\Pi^{\alpha\beta\gamma}$ is the conjugate momentum of the 2-form $B_{\mu\nu}$. Upon variation with respect to the conjugate momentum we obtain
\beq\label{mom2form}
\Pi_{\alpha\beta\gamma}=\partial_{[\alpha} B_{\beta\gamma]}
\eeq
while the 2-form field equations give
\beq
\partial_\alpha \Pi^{\alpha\beta\gamma}=0.
\eeq
These are of course nothing but the Hamilton equations of a Kalb-Rammond field. Since the conjugate momentum is divergence-free as imposed by the 2-form field equation, in 4 dimensions,z it can be solved by $H^{\alpha\beta\gamma}\propto\epsilon^{\alpha\beta\gamma\mu}\partial_\mu\phi$, with $\phi$ a scalar field. We can then plug the solution of the constraint imposed by the 2-form into the action so we end up with
\beq
\cS=\int\dd^4x\sqrt{-g}\partial_\mu\phi\partial^\mu\phi,
\eeq
showing that a Kalb-Rammond field indeed propagates a scalar degree of freedom. This procedure can formally be done at the level of the path integral $Z=\int [D\Pi] [DB] e^{i\cS}$ by integrating out the 2-form field. Since the action is a quadratic form in the momentum, the integration can straightforwardly be performed  giving a functional delta that imposes the constraint.

Let us now turn to the case of a massive 2-form field. Again, we will resort to its first order formulation
\beq
\cS=-\frac16\int\dd^4x\sqrt{-g}\left(\Pi^{\alpha\beta\gamma}\partial_{[\alpha} B_{\beta\gamma]}-\frac12\Pi^2+{\frac{1}{2}}m^2B^2\right),
\eeq
where we see that the 2-form now becomes an auxiliary field instead of a Lagrange multiplier imposing the divergence-free constraint on the conjugate momentum as in the massless case. The field equation for the momentum again gives its relation with the derivatives of the 2-form as \eqref{mom2form}. The 2-form equations however now give
\beq
\partial_\alpha \Pi^{\alpha\beta\gamma}- m^2 B^{\be\ga}=0.
\eeq
Using this equation to solve for the 2-form   in terms of the conjugate momentum and plugging it into the action (notice that this is an algebraic equation for $B_{\mu\nu}$), we can rewrite the action as
\beq
\cS=\frac{1}{12}\int\dd^4x\sqrt{-g}\left({\frac{1}{m^2}}\partial_\alpha\Pi^{\alpha\beta\gamma}\partial^{\lambda}\Pi_{\lambda\beta\gamma}+\Pi^2\right).
\eeq
We can now dualise this action by means of $H^{\alpha\beta\gamma}\propto \epsilon^{\alpha\beta\gamma\mu}A_\mu$ and after canonically normalising $A_\mu$ by $A_\mu\mapsto (m/2) A_\mu$, the above action can equivalently be expressed as
\beq
\cS=\int\dd^4x\sqrt{-g}\left(-\frac14 F_{\mu\nu}F^{\mu\nu}-\frac12 M^2 A^2\right),
\eeq
with $M^2=3m^2$. We have then that our original massive 2-form field is dual to a Proca theory and, consequently, it propagates three degrees of freedom.\\ 

For completeness, let us also discuss what happens for more general interacting 2-form fields. Again, starting from its first order formulation, we can write 
\beq
\cS=\int\dd^4x\sqrt{-g}\Big[\Pi^{\alpha\beta\gamma}\partial_{[\alpha} B_{\beta\gamma]}-\mathcal{H}(B_{\mu\nu},\Pi^{\alpha\beta\gamma})\Big],
\eeq
where $\mathcal{H}$ is the Hamilton function that defines the interacting theory. Following the same procedure, we can obtain the 2-form field equations as
\beq
\partial_\alpha \Pi^{\alpha\mu\nu}=-\frac{\partial\mathcal{H}}{\partial B_{\mu\nu}}.
\label{Eq:eqdH}
\eeq
This equation now can be algebraically solved (at least formally) for the 2-form field to obtain $B^{\mu\nu}=B^{\mu\nu}(\Pi,\partial\cdot\Pi)$. We can then integrate out the 2-form field by plugging this solution into the action so we obtain $\cS=\cS[\Pi,\partial\cdot\Pi]$. If we dualise this theory to a vector field as above, we finally get that our original action can be rewritten as $\cS=\cS[F_{\mu\nu},A_\alpha]$, i.e., as an interacting massive vector field. If the Hamiltonian function does not explicitly depend on the 2-form field, i.e., we have a gauge 2-form field, then Eq. (\ref{Eq:eqdH}) is instead solved by $H^{\alpha\beta\gamma}=\epsilon^{\alpha\beta\gamma\mu}\partial_\mu\phi$ as in the massless case above, so the action can instead be expressed as $\cS=\cS[(\partial\phi)^2]$ that describes an interacting shift-symmetric scalar field. Notice that this dualisation procedure can also be applied to cases when the 2-form field is coupled to some matter fields and even for non-Abelian 2-form fields with some internal group structure.

\vspace{-0.4cm}
\section{A spin-1 proxy}
In this Appendix we will clarify some potentially confusing issues regarding the counting of degrees of freedom and the origin of the projective ghost. For that, we will use the following simpler lower spin Lagrangian:
\beq
\cL=-\frac14 F_{\mu\nu}F^{\mu\nu}+\lambda A^\mu\partial_\mu\varphi+\frac12m^2A^2.
\eeq
In this Lagrangian, the spin-1 and the scalar fields play the roles of the 2-form and the projective mode respectively. We see that the scalar imposes the constraint
\beq
\partial_\mu A^\mu=0,
\eeq
while the spin-1 field equations are
\beq
\partial_\mu F^{\mu\nu}+\lambda\partial^\nu\varphi + m^2A^\nu=0.
\eeq
Let us start by counting degrees of freedom. The expected number is the 3 polarisations of the massive vector field plus 1 associated to the scalar, giving a total of 4. This counting is indeed correct, but one may worry that the constraint imposed by the scalar field could alter this counting. A direct counting can be carried out by clarifying the number of initial Cauchy data one needs to provide. In principle we should give the values of all the dynamical fields and their first derivatives. However, we have constraints so some of them can actually be expressed in terms of the remaining ones. The scalar field equation gives
\beq
\dot{A}_0+\partial_i A^i=0
\eeq
so we see that the derivative of $A_0$ on the Cauchy surface is determined by the initial values of $A^i$. On the other hand, the temporal component of the vector field equations gives
\beq
-\partial_i\dot{A}^i+\lambda\dot{\varphi} + m^2A^0=0
\eeq
that allows to express the initial derivative of the scalar field in terms of $A_0$ and $\dot{A}^i$ on the Cauchy surface. We have exhausted all the constraints and we obtain that we only need to give the initial values of $A^i$, $\dot{A}^i$, $\varphi$ and $A_0$, what corresponds to 8 phase space conditions, i.e., there are 4 dynamical degrees of freedom, in agreement with our expectation.\\

An alternative way of counting the number of propagating modes is by resorting to the Stueckelberg trick and taking the decoupling limit, as we have done for the 2-form case as well. If we restore the $U(1)$ gauge-invariance by replacing $A_\mu\rightarrow A_\mu+\frac{1}{m}\partial_\mu\chi$, the Lagrangian becomes
\beq
\cL=-\frac14F^2+\partial^\mu\chi\partial_\mu\bar{\varphi}+\frac12\partial^\mu\chi\partial_\mu\chi+\frac12m^2A^2+mA^\mu(\partial_\mu\bar{\varphi}+\partial_\mu\chi),
\eeq
where $\bar{\varphi}=\frac{\lambda}{m}\varphi$. If we now take the decoupling limit $m\rightarrow0$, the Lagrangian decouples as
\beq
\cL=-\frac14F^2+\partial^\mu\chi\partial_\mu\bar{\varphi}+\frac12\partial^\mu\chi\partial_\mu\chi,
\eeq
where we have isolated the longitudinal mode of the vector from the transverse modes. In this limit, it becomes much more apparent that the theory progates 4 degrees of freedom corresponding to the 2 transverse modes, the longitudinal polarisation and the scalar field. Furthermore, this limit also allows to clearly see the pathological behaviour of the scalar field because of the absence of a term $(\partial\bar{\varphi})^2$. This can be seen by noticing that the eigenvalues of the kinetic matrix are $+1$ and $-1$. The scalar sector can be diagonalised by means of $\chi\rightarrow \chi-\bar{\varphi}$ so we have
\beq
\cL=-\frac14F^2-\frac12\partial^\mu\bar{\varphi}\partial_\mu\bar{\varphi}+\frac12\partial^\mu\chi\partial_\mu\chi,
\eeq
where we see the unavoidable presence of a ghost.

\newpage

%\bibliography{Bibliography}
%\bibliographystyle{unsrt}
%\bibliographystyle{JHEP}
\bibliography{Bibliography}

\end{document}